\documentclass[pre,twocolumn,nobalancelastpage,amsmath,amssymb,showpacs,
nofootinbib, superscriptaddress]{revtex4-1}

\usepackage{graphics}      
\usepackage{graphicx}      
\usepackage{longtable}     
\usepackage{url}           
\usepackage{bm}            
\usepackage{xcolor}
\usepackage{amsmath}
\usepackage{dcolumn}
\usepackage{mathtools}
\usepackage{hyperref}

\begin{document}
\title{Dynamic length scales in athermal, shear-driven, jamming of frictionless disks in two dimensions}
\author{Peter Olsson}
\affiliation{Department of Physics, Ume{\aa} University, 901 87 Ume{\aa}, Sweden}
\author{S. Teitel}
\affiliation{Department of Physics and Astronomy, University of Rochester, Rochester, NY 14627}
\date{\today}

\begin{abstract}
We carry out numerical simulations of athermally sheared, bidisperse, frictionless disks in two dimensions.  From an appropriately defined velocity correlation function, we determine that there are two diverging length scales, $\xi$ and $\ell$, as the jamming transition is approached.  We analyze our results using a critical scaling ansatz for the correlation function, and argue that the more divergent length $\ell$ is a consequence of a dangerous irrelevant scaling variable, and that it is $\xi$ which is the correlation length that determines the divergence of the system viscosity as jamming is approached from below in the liquid phase.  We find that $\xi\sim (\phi_J-\phi)^{-\nu}$ diverges with the critical exponent $\nu=1$.  We provide evidence that  $\xi$ measures the length scale of fluctuations in the rotation of the particle velocity field, while $\ell$ measures the length scale of fluctuations in the divergence of the velocity field.
\end{abstract}
\maketitle

\section{Introduction}
\label{sec:intro}

Athermal granular and related soft matter materials, such as non-Brownian suspensions, emulsions, and foams, all undergo a phase transition from a liquid-like state to a rigid but disordered solid state as the packing fraction $\phi$ of the granular particles increases.  This is known as the jamming transition \cite{LiuNagel,OHern}.  In isotropic  jamming, mechanically stable configurations are generated by  isotropically compressing the system, or by quenching random initial configurations \cite{OHern,Wyart,Chaudhuri,Vagberg.PRE.2011}.  At low $\phi$ particles may avoid each other and the system pressure $p=0$.  At a critical $\phi_J$ a system spanning rigid cluster forms and the system pressure becomes finite.  In shear-driven jamming \cite{OT1,OT2,VagbergOlssonTeitel,OT3, Hatano1,Hatano2,Hatano3,Otsuki,Heussinger1,Heussinger2}  the system is uniformly sheared at a fixed strain rate $\dot\gamma$.  In a system with a Newtonian rheology, such as for particles in suspension, at low $\phi$ and small $\dot\gamma$ the system flows with a shear stress $\sigma\propto \dot\gamma$, and thus a finite viscosity $\eta=\sigma/\dot\gamma$.  But above a critical $\phi_J$ the system develops a finite yield stress and $\lim_{\dot\gamma\to 0}\sigma = \sigma_0(\phi)>0$.

For the idealized case of frictionless particles, the jamming transitions behave in many respects like continuous phase transitions \cite{OHern,OT1,OT2,VagbergOlssonTeitel}.  In isotropic jamming, in the limit of quasistatic compression, the pressure $p(\phi)$  increases algebraically from zero as $\phi$ increases above $\phi_J$ \cite{OHern}.  In shear-driven jamming, in the limit $\dot\gamma\to 0$ of quasistatic shearing, the yield stress $\sigma_0(\phi)$ (as well as the pressure $p_0(\phi)$ along the yield stress line) similarly increases algebraically from zero as $\phi$ increases above $\phi_J$ \cite{OT1,OT2,Heussinger1}.   This analogy with continuous phase transitions suggests that there should be a critical correlation length  $\xi$ that diverges as the jamming transition is approached, and it is the divergence of this $\xi$ that sets the singular behavior of other observable quantities.  The goal of this work is to identify this $\xi$ for shear-driven jamming in a simple model  with Newtonian rheology in two dimensions, and to determine the critical exponent $\nu$ that controls it algebraic divergence at $\phi_J$.

For isotropic jamming, an analysis of the modes of small vibration of  mechanically stable packings above $\phi_J$  by Silbert et al. \cite{Silbert}
led to diverging longitudinal and transverse lengths in the solid phase, $\xi_L\sim (\phi-\phi_J)^{-\nu_L}$ and $\xi_T\sim(\phi-\phi_J)^{-\nu_T}$, with $\nu_L\approx 1/2$ and $\nu_T\approx 1/4$.  A diverging isostatic length scale  $\ell^*\sim 1/(z-z_\mathrm{iso})$, measuring the deviation of the average particle contact number $z$ from  the frictionless isostatic value $z_\mathrm{iso}=2d$ in $d$ dimensions, was predicted from analytical arguments by Wyart et al. \cite{Wyart2}.  Since $z-z_\mathrm{iso}\sim (\phi-\phi_J)^{1/2}$ \cite{OHern}, one finds $\ell^*\sim\xi_L$.  Recently, Hexner et al. \cite{Hexner} have proposed two new diverging length scales above $\phi_J$, $\xi_z$ associated with correlations of the average particle contact number, and $\xi_f$ associated with contact number fluctuations.  They find $\xi_z\sim 1/(z-z_\mathrm{iso})^{\nu_z}$, with $\nu_z= 0.7$ in $d=2$ dimensions and 0.85 in $d=3$ dimensions; and $\xi_f\sim1/(z-z_\mathrm{iso})^{\nu_f}$, with $\nu_f=1.07$ in $d=2$ and 1.29 in $d=3$.

Drocco et al. \cite{Drocco} determined a diverging length scale with exponent $\nu=0.6 - 0.7$ by considering the size of the cluster of particles that is dragged along by an intruder forced through the system at different packings $\phi$ in two dimensions.  A similar value of $\nu$ was found by O'Hern et al. \cite{OHern} from looking at the scaling of the  critical $\phi_J$ with system size in mechanically stable packings in both two and three dimensions.  However a finite-size scaling analysis by V{\aa}gberg et al., \cite{VOT} for both isotropic and shear-driven jamming in two dimensions, argued that the value $\nu\approx 0.7$  was an artifact of not including corrections to scaling, and that once such corrections are included one finds $\nu\approx 1$.  However others have challenged whether such a finite-size scaling analysis correctly probes the correlation length in $d=2$ dimensions.  Above the upper critical dimension ($ucd$), where mean-field results hold, one expects quantities to scale with system length $L$ according to $L^{d/2}$.  Since some arguments suggest that $ucd=2$ for the jamming transition \cite{Wyart,Wyart3,Goodrich,Charbonneau,Goodrich2}, a value $\nu=1$, determined from finite-size scaling in $d=2$, could thus reflect this mean-field behavior rather than the scaling of the correlation length $\xi$.

For sheared systems, Heussinger and Barrat \cite{Heussinger1} argued that, for $\phi<\phi_J$, one could define an isostatic length scale $\ell^*\sim 1/(z_\mathrm{iso}-z)$, similarly to that defined above $\phi_J$.  However, unlike for mechanically stable states above $\phi_J$, they found for quasistatically sheared configurations below jamming that $z_\mathrm{iso}-z\sim (\phi_J-\phi)$, thus implying $\ell^*\sim (\phi_J-\phi)^{-\nu}$ with $\nu=1$.
{\color{black}Numerical results in this work then found $\nu$ in the range of 0.8--1.0.}
In later work, Heussinger et al. \cite{Heussinger3} defined several other length scales for sheared systems below jamming, $\phi<\phi_J$, obtained by measuring the variation of different properties as the system evolves with increasing strain $\gamma$.
From measurements of the mean-squared particle displacement $\Delta$  they found $\ell_\Delta\sim(\phi_J-\phi)^{-1.1}$; from measurements of particle overlap $Q$  they found $\ell_Q\sim (\phi_J-\phi)^{-0.9}$; and from measurements of the dynamical susceptibility $\chi_4$ they found $\xi_4\sim(\phi_J-\phi)^{-0.9}$. 
These observations could be consistent with $\nu=1$.  D{\"u}ring et al. \cite{During2}, however, predicted that for sheared systems there are two diverging length scales, $\ell_c\sim1/\sqrt{z_\mathrm{iso}-z}$ and $\ell_r\sim \sqrt{p/\dot\gamma}$, but argue that it is $\ell_c$ that sets the length scale of velocity correlations.  Assuming $z_\mathrm{iso}-z\sim (\phi_J-\phi)$ then gives $\nu=1/2$.
While Ref.~\cite{During2} presents numerical support for these two lengths scales for a model network of rigid rods, the numerical results for particle suspensions are less clear.   For suspensions, the authors state that their system sizes are not sufficiently large for them to numerically test their prediction concerning $\ell_r$. The smaller length $\ell_c$  describes only the decay of the velocity correlation function on relatively short length scales $r\lesssim \ell_c$, such that the correlation decays to a relative value of about 0.2 (see their Fig.~7).  In contrast, here we will be concerned with behavior on long length scales, where the correlation has already decayed to a relative value of around 0.05 and smaller.

In this work we reexamine the question of the correlation length in athermally sheared suspensions of frictionless particles.  Our goal is to make a direct 
measurement of the correlation length by looking at the spatial decay of an appropriate velocity correlation function.
Experimental measurements by Pouliquen \cite{Pouliquen}, of velocity correlations for grains flowing down a rough inclined plane, showed an increase in the correlation length as the angle of incline, and hence the average flow velocity, decreased.  
In an early work on shear-driven jamming \cite{OT1} we proposed a correlation length $\xi$, obtained from the transverse velocity correlation function.  Carrying out a critical scaling analysis we concluded that the correlation length exponent was $\nu\approx 0.6$, similarly to the value obtained in several earlier works \cite{OHern,Drocco}. In this work, however, we argue that our previous analysis was incorrect, because of a failure to appreciate the effects of multiple length scales.  We present a new, more careful, analysis of a somewhat different velocity correlation function, and now conclude that $\nu=1$ for two dimensions.

The remainder of our paper is organized as follows.  In Sec.~\ref{model} we present the model we use for our simulations, which is the Durian ``mean-field" model for foams \cite{Durian}.  In Sec.~\ref{oldC} we review our earlier results \cite{OT1} for the transverse velocity correlation function and indicate the difficulty with our earlier analysis.  In Sec.~\ref{newC} we define a new velocity correlation function and present our numerical analysis which demonstrates that there are two different divergent lengths, $\xi$ and $\ell$, each diverging with a different critical exponent at jamming.  In Sec.~\ref{scale} we present a scaling analysis for our velocity correlation and  argue that the diverging length $\ell$ is a consequence of a dangerous irrelevant scaling variable.  In Sec.~\ref{eta} we show that, while $\ell$ diverges more rapidly than $\xi$ as jamming is approached,  it is nonetheless $\xi$ that is the proper correlation length that determines the divergence of the viscosity as jamming is approached.
In Sec.~\ref{vorticity}  we provide a physical interpretation for the two diverging lengths, showing that $\xi$ is associated with the rotation of the particles' velocity field, while $\ell$  is associated with the divergence of the velocity field.  Finally in Sec.~\ref{sum} we summarize our conclusions.

\section{Model}
\label{model}

We use a well studied model of size-bidisperse, frictionless, soft-core circular disks in two dimensions \cite{OHern}.  We take equal numbers of big and small disks with diameter ratio of $d_b/d_s=1.4$.  Particle center of mass positions and velocities are denoted as $\mathbf{r}_i$ and $\mathbf{v}_i=d\mathbf{r}_i/dt$, respectively.

When two particles overlap, they experience a repulsive elastic force.
If $r_{ij}=|\mathbf{r}_i-\mathbf{r}_j|$ is the center-to-center distance between two disks, then a pair of disks will overlap whenever $r_{ij}<d_{ij}=(d_i+d_j)/2$.  The interaction  between particles is taken as simple one-sided harmonic potential,
\begin{equation}
V(r_{ij})=\left\{
\begin{array}{cc}
\frac{1}{2}k_e(1-r_{ij}/d_{ij})^2, & r_{ij}<d_{ij}\\
0, &r_{ij}>d_{ij}
\end{array}
\right.
\end{equation}
where $k_e$ is the  stiffness constant of the interaction.
When particles overlap, the elastic force on particle $i$ due to its contact with $j$ is thus 
\begin{equation}
\mathbf{F}_{ij}^\mathrm{el}=-\frac{dV(r_{ij})}{d\mathbf{r}_i}=k_e(1-r_{ij}/d_{ij})\mathbf{\hat n}_{ij}
\end{equation}
with $\mathbf{\hat n}_{ij}$ the inward  normal to the surface of particle $i$ at its point of contact with $j$. 

Particles also experience a dissipative force, which we model as a viscous drag with respect to a uniformly sheared host medium, as for a particle in suspension,
\begin{equation}
\mathbf{F}_i^\mathrm{dis} = -k_d[\mathbf{v}_i-\dot\gamma y_i\mathbf{\hat x}]
\end{equation}
where $\dot\gamma$ is the shear strain rate, $k_d$ the viscous drag coefficient, and the flow is in the $\mathbf{\hat x}$ direction \cite{OT1,OT2}.  

We  use an overdamped equation of motion, 
\begin{equation}
{\sum_j}^\prime \mathbf{F}_{ij}^\mathrm{el} + \mathbf{F}_i^\mathrm{dis}=0
\end{equation}
where the sum is over all particles $j$ in contact with $i$.  This leads to the equation of motion for particle $i$,
\begin{equation}
\frac{d\mathbf{r}_i}{dt}=\dot\gamma y_i\mathbf{\hat x} +\frac{1}{k_d}{\sum_j}^\prime \mathbf{F}_{ij}^\mathrm{el},
\end{equation}
which is equivalent to the Durian bubble model \cite{Durian} for foams in his ``mean field" limit.  Uniform simple shearing is applied using Lees-Edwards boundary conditions \cite{LeesEdwards} on a system of equal length and height $L$.  While this model is greatly simplified, it is a well studied and commonly used model for studying the criticality of frictionless shear-driven jamming for a system with Newtonian rheology \cite{Tewari,Andreotti,Lerner,Vagberg.PRL.2014,DeGiuli,Berthier}.

For our simulations we  take the unit of length as $d_s=1$, the unit of energy as $k_e=1$, and the unit of time as $t_0= k_d d_s^2/k_e=1$.  The equations of motion are integrated using the Heun method with an integration step $\Delta t =0.2 t_0$. Unless stated otherwise,  our simulations  use a total of $N=262144$ particles, varying $\dot\gamma$ at the fixed packing $\phi=0.8433$, which we have previously determined \cite{OT2,VOT} to be the shear-driven jamming $\phi_J$ of our model.  At this packing the system has a length $L\approx 601$.  We typically shear our simulations to a total strain of $\gamma_\mathrm{tot}\sim O(10^3)$ for the largest $\dot\gamma$, and to $\dot\gamma_\mathrm{tot}\sim O(5)$ for the smallest $\dot\gamma$.  We start our shearing runs from an initial random configuration at the largest $\dot\gamma$.  For each smaller $\dot\gamma$ we start with a configuration from the next larger $\dot\gamma$.

\section{Transverse velocity correlation}
\label{oldC}

{\color{black}In our previous work on shear-driven jamming \cite{OT1} we proposed a measure of the correlation length from consideration of the transverse velocity correlation function, and argued for a correlation length exponent $\nu\approx 0.6$.  In this section we  show why this prior work is incorrect.  However, first we make a more general comment about our prior work \cite{OT1}.  While we believe that the multivariable critical scaling introduced in \cite{OT1} is valid and provided a new understanding of the shear-driven jamming transition, none of the specific numerical values for critical exponents or the jamming density that we reported in Ref.~\cite{OT1} are, to our current understanding, correct.  The scaling collapses that determined the values of critical parameters in Ref.~\cite{OT1} were obtained by eyeball estimates of goodness of fit.  We have since demonstrated (see Ref.~\cite{VagbergOlssonTeitel} Sec.~V.A) that such eyeball estimates can often be misleading and that a more systematic analysis is required.  We carried out such a systematic study for our current model in Ref.~\cite{OT2}, where we found that corrections to scaling (ignored in Ref.~\cite{OT1}) were needed to correctly describe our numerical results.  We will show in the present work that corrections to scaling are similarly needed for a correct description of the velocity correlation function.  Note that the values of $\phi_J$ and the critical exponent $1/z\nu$ that we use in the present analysis are the values obtained by us in Ref.~\cite{OT2}. We now turn  back to the velocity correlation function.}

Asymptotically close to the jamming point the critical scaling equation for the correlation length, ignoring corrections to scaling, is
\begin{equation}
\xi(\phi,\dot\gamma)=b h(\delta\phi b^{1/\nu}, \dot\gamma b^z),
\label{scalexi}
\end{equation}
with $\delta\phi=\phi-\phi_J$, $\nu$ is the correlation critical exponent, $z$ the dynamic critical exponent, and $b$ an arbitrary length rescaling factor \cite{VagbergOlssonTeitel}.  If we set $b=|\delta\phi|^{-\nu}$, then the above becomes 
\begin{equation}
\xi=|\delta\phi|^{-\nu}h(\pm1,\dot\gamma/|\delta\phi|^{z\nu})
\label{xi1}
\end{equation}
where $+1$ is for $\phi>\phi_J$ and $-1$ is for $\phi<\phi_J$.  Taking $\dot\gamma\to 0$, we expect $h(\pm 1,0)$ is a finite constant, and we then have the usual $\xi\sim|\delta\phi|^{-\nu}$.  But if we take $b=\dot\gamma^{-1/z}$, then  we  get
\begin{equation}
\xi = \dot\gamma^{-1/z}h(\delta\phi/\dot\gamma^{1/z\nu},1).
\label{xi2}
\end{equation}
Expecting $h(0,1)$ to be a finite constant, at $\phi=\phi_J$ ($\delta\phi=0$) we then get $\xi\sim\dot\gamma^{-1/z}$.

In this section and the next, we  consider the behavior of the velocity correlation for varying $\dot\gamma$ at fixed $\phi=\phi_J$, and thus attempt to determine the critical exponent $z$.  Using the value $1/z\nu=0.26\pm0.02$, obtained from our earlier scaling analysis of the stress \cite{OT2}, we can then find the value of $\nu$.

In our earlier work \cite{OT1} we considered the correlations of the component of the particle velocity transverse to the direction of the shear flow,
\begin{equation}
g_y(x)=\langle v_y(\mathbf{r}_0)v_y(\mathbf{r}_0+x\mathbf{\hat x})\rangle /\langle v_y^2\rangle.
\end{equation}
The normalization is chosen so that $g_y(0)=1$.
By translational symmetry, the above correlation is independent of the position $\mathbf{r}_0$, and depends only on the separation $x\mathbf{\hat x}$.
To compute velocity correlations we use the following method.  The continuous system is discretized by a square grid of boxes, where the grid box is sufficiently small that only a single particle can have its center in any given box.  The boxes that contain the center of a particle are then assigned the velocity of that particle.  The correlation is then computed by averaging over pairs of boxes with the specified separation, and then averaging over different configurations within the sheared steady-state ensemble.  Empty boxes are not included in this calculation.

\begin{figure}
\centering
\includegraphics[width=3.3in]{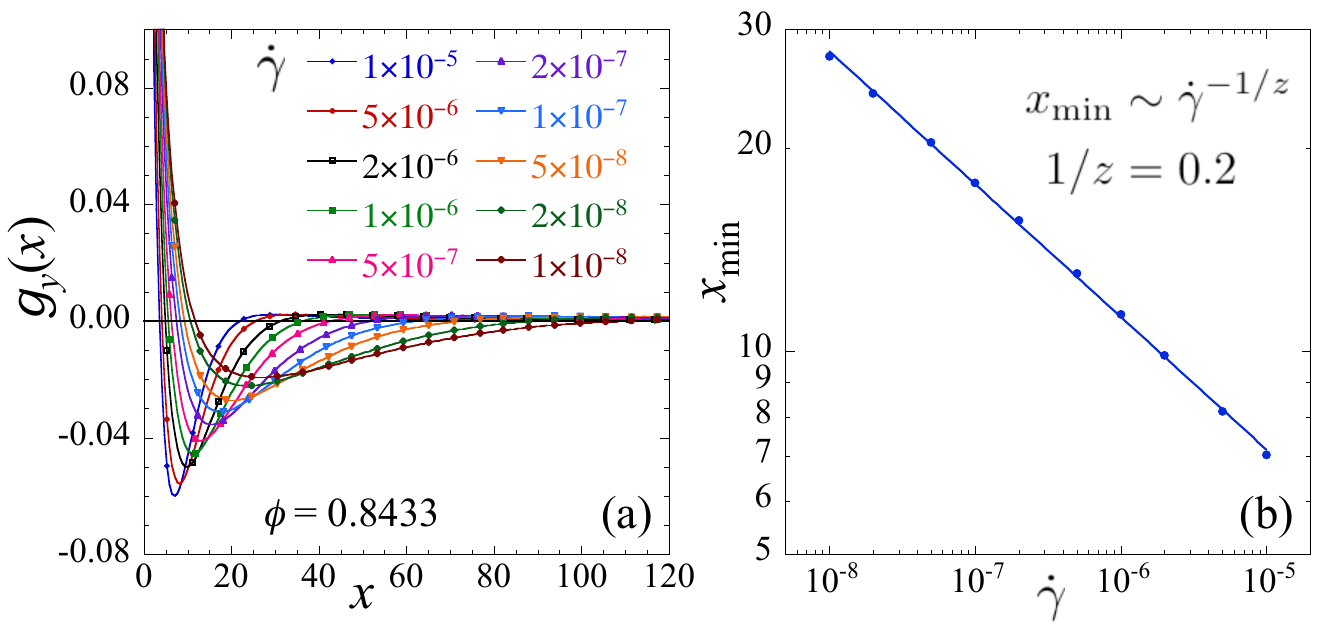}
\caption{(a) Transverse velocity correlation $g_y(x)$ vs $x$ for different strain rates $\dot\gamma$ at fixed $\phi=0.8433\approx \phi_J$. Symbols are shown only on a small subset of the data points to help identify the different curves.  (b) Location $x_\mathrm{min}$ of the minimum in $g_y(x)$ vs $\dot\gamma$ for fixed $\phi=0.8433$.  The straight line is the fit $x_\mathrm{min}\sim\dot\gamma^{-z}$ with $1/z=0.2$.
}
\label{gy-x} 
\end{figure}

In Fig.~\ref{gy-x}(a) we plot $g_y(x)$ vs $x$ for different strain rates $\dot\gamma$ at the fixed $\phi=0.8433\approx\phi_J$ \cite{OT2}.  We see that the correlation decreases, reaches a minimum at a distance we will denote as $x_\mathrm{min}$, and then increases again to decay to zero.  The location of the minimum $x_\mathrm{min}$ sets a length scale for the system.  We see that $x_\mathrm{min}$ increases as $\dot\gamma$ decreases and one approaches the jamming critical point.  In Fig.~\ref{gy-x}(b) we plot $x_\mathrm{min}$ vs $\dot\gamma$ and find a reasonable fit to the powerlaw $x_\mathrm{min}\sim \dot\gamma^{-1/z}$, with $1/z=0.2$.  Using the value $1/z\nu=0.26$, obtained from our earlier analysis of stress \cite{OT2}, we then  get a value of $\nu=0.77$.  This value differs from the $\nu\approx 0.6$ of our earlier work \cite{OT1} because the value of the critical $\phi_J$ claimed in that work was later found, by a more careful analysis \cite{OT2}, to be too low.  However the analysis presented here still illustrates how one typically gets values $\nu< 1$ from such an approach.

\begin{figure}
\centering
\includegraphics[width=3.3in]{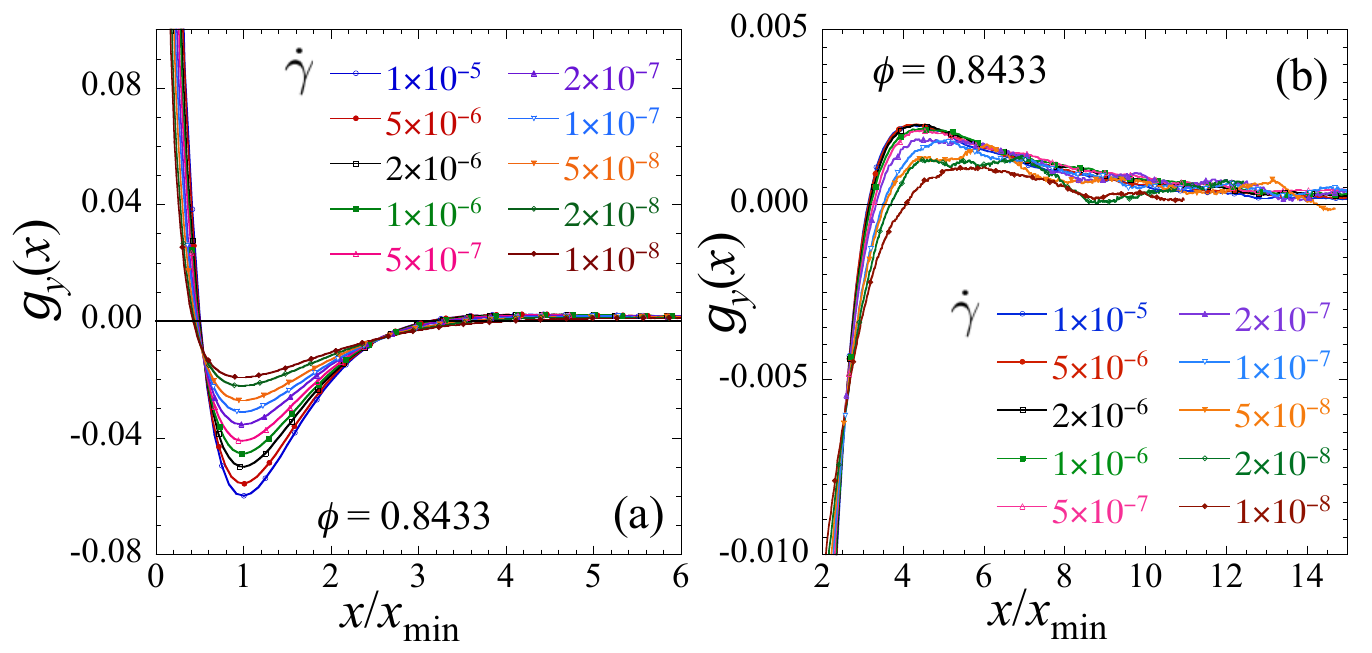}
\caption{(a) Transverse velocity correlation $g_y(x)$ vs scaled distance $x/x_\mathrm{min}$ for different strain rates $\dot\gamma$ at fixed $\phi=0.8433\approx \phi_J$. (b) A blow up of the plot in (a) focusing on the region above $x_\mathrm{min}$ where $g_y(x)$ becomes positive.  Symbols are shown only on a small subset of the data points to help identify the different curves.  
}
\label{gy-x-scaled} 
\end{figure}

Although the fit in Fig.~\ref{gy-x}(b) looks good, there are problems with this analysis.  
If $x_\mathrm{min}$ were indeed the correlation length, and this were the only important length scale in the problem for distances $x>x_\mathrm{min}$, then we would expect that all features in the curves $g_y(x)$ vs $x$ for different $\dot\gamma$ would align when  plotting $g_y(x)$ vs $x/x_\mathrm{min}$.  We show such a scaled plot in Fig.~\ref{gy-x-scaled}.  In Fig.~\ref{gy-x-scaled}(a) we show a range of $x/x_\mathrm{min}$ that includes the minimum at $x/x_\mathrm{min}=1$.  In Fig.~\ref{gy-x-scaled}(b) we show a blow up of the plot in Fig.~\ref{gy-x-scaled}(a) that focuses on the region above the minimum.  Here we clearly see that $g_y(x)$ does not stay negative as it decays to zero, but in fact turns positive again, reaches a maximum, and then decays to zero.  Comparing the curves of different $\dot\gamma$, neither the location $x_0>x_\mathrm{min}$ where $g_y(x)$ crosses zero, nor the location $x_\mathrm{max}$ where $g_y(x)$ has its maximum, align when plotting vs $x/x_\mathrm{min}$.  Both $x_0/x_\mathrm{min}$ and $x_\mathrm{max}/x_\mathrm{min}$ increase as $\dot\gamma$ decreases.  This thus indicates that the correlation $g_y(x)$ at  large $x>x_\mathrm{min}$ is governed by more than one  length scale.  

While the divergence of $x_\mathrm{min}$ as jamming is approached indicates that there is indeed a diverging length scale in the problem, it would seem that the value of $x_\mathrm{min}$ is determined by the competition of two or more length scales that diverge differently.  The exponent $1/z$ found in Fig.~\ref{gy-x}(b) from the fit of $x_\mathrm{min}\sim\dot\gamma^{-1/z}$ should be regarded as only an effective exponent for a specific range of $\dot\gamma$, rather than the true dynamic exponent associated with the divergence of the correlation length $\xi$ as one gets asymptotically close to jamming.

One could attempt to determine the different length scales contributing to $g_y(x)$ if one had a good analytic approximation to the functional form of $g_y(x)$.  Fitting to that form would allow one to extract the different lengths and see how they separately behave as jamming is approached.  However, the complex structure of $g_y(x)$, decreasing, then increasing, and then decreasing again toward zero, leaves us without any good analytical form for such a fit.  In the following section, we therefore consider an alternative velocity correlation for which such an analysis is possible.


\section{Alternative velocity correlation}
\label{newC}

To measure a correlation length, one would in principle like to find a quantity whose correlation displays a simple exponential decay at large lengths.  Clearly $g_y(x)$ does not do so.  We have also considered
\begin{equation}
g_x(x)=\langle\delta v_x(\mathbf{r}_0)\delta v_x (\mathbf{r}_0+x\mathbf{\hat x})\rangle/\langle \delta v_x^2\rangle
\end{equation}
where
\begin{equation}
\delta\mathbf{v}_i=\mathbf{v}_i-\dot\gamma y_i\mathbf{\hat x}
\end{equation}
is the nonaffine part of the velocity of particle $i$, i.e., the fluctuation of the velocity away from a uniform shear flow (note $\delta v_y = v_y$ since the affine part of the velocity is strictly in the $\mathbf{\hat x}$ direction).
In general, $g_x(x)$ is also not a simple exponential decay, but unlike $g_y(x)$ it appears to have only a single extremum; as $x$ increases, $g_x(x)$ decreases, reaches a minimum, and then increases to decay to zero while staying negative.  We thus find that we can reasonably parametrize $g_x(x)$ as the sum of two exponentials with possibly different decay lengths.

While such a procedure works reasonably well for $g_x(x)$, after some trial and error, we have found that a two exponential parametrization  works even better \cite{gxvsg}, giving more accurate results, when applied to an alternative velocity correlation function given by
\begin{equation}
g(x) = \dfrac{\langle \delta v_x(\mathbf{r}_0)\delta v_x(\mathbf{r}_0+x\mathbf{\hat x})\rangle
-\langle \delta v_y(\mathbf{r}_0)\delta v_y(\mathbf{r}_0+x\mathbf{\hat x})\rangle}{\langle|\delta\mathbf{v}|^2\rangle/2}.
\label{gmix}
\end{equation}
Hence, in this section we will focus on  $g(x)$.  We will provide a physical interpretation for this particular correlation later in Sec.~\ref{vorticity}.
Defining
\begin{equation}
f(x)=A\mathrm{e}^{-x/\xi} - B\mathrm{e}^{-x/\ell},
\label{exp2}
\end{equation}
with $A$ and $B$ both positive,
 we find a reasonable fit to  Eq.~(\ref{gmix})  by taking
\begin{equation}
g(x)=f(x)+f(L-x),
\label{exp2pbc}
\end{equation}
where the second term  is used to enforce the periodic boundary condition, $g(x)=g(L-x)$.

\begin{figure}
\centering
\includegraphics[width=3.3in]{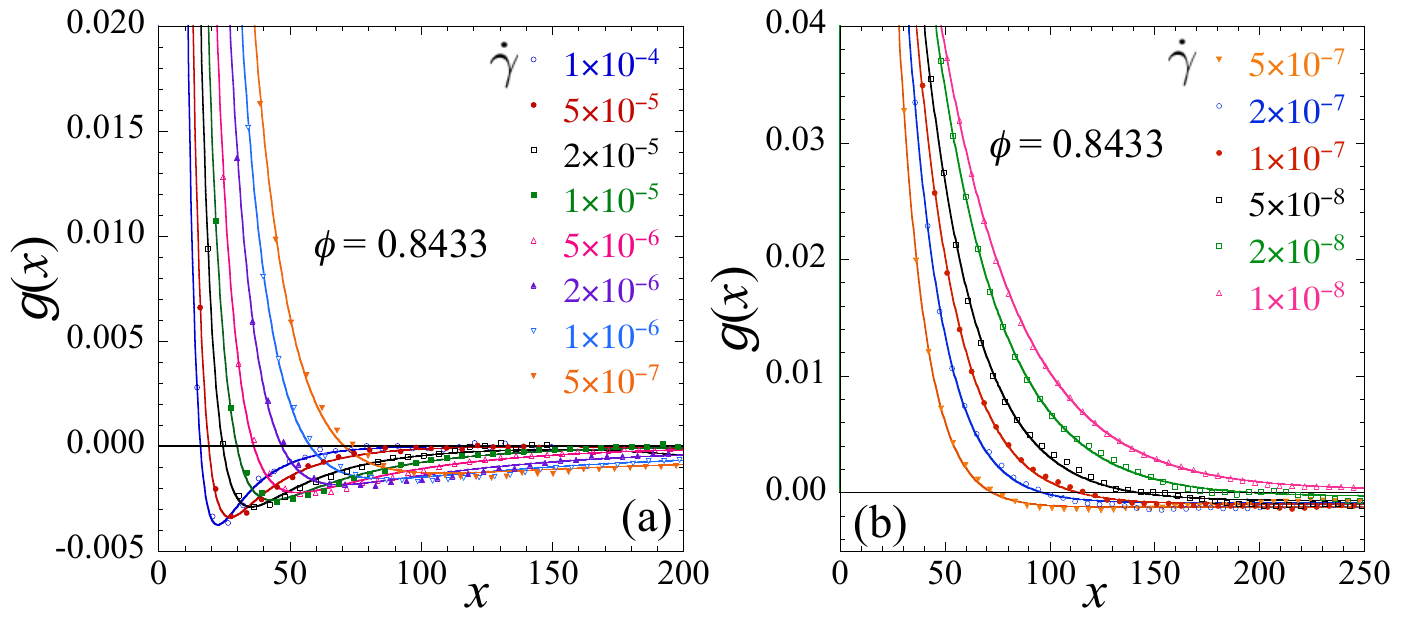}
\caption{Velocity correlation $g(x)$ of Eq.~(\ref{gmix}) vs $x$ for different strain rates $\dot\gamma$ at fixed $\phi=0.8433\approx \phi_J$. 
(a) Correlation for large strain rates $\dot\gamma\ge 5\times 10^{-7}$, where $g(x)$ is nonmonotonic.  Solid lines are fits to the form of Eq.~(\ref{exp2}).  (b) Correlation for small strain rates $\dot\gamma\le 5\times 10^{-7}$, where $g(x)$ monotonically decreases.  Solid lines are fits to the form of Eq.~(\ref{exp1}).  In both (a) and (b) we show only every 10th data point at each $\dot\gamma$, for the sake of clarity.  The system has $N=262144$ particles for $\dot\gamma\le10^{-5}$, and $N=65536$ particles for $\dot\gamma>10^{-5}$.
}
\label{g-x} 
\end{figure}

In Fig.~\ref{g-x} we plot $g(x)$ vs $x$ for different strain rates $\dot\gamma$ at the fixed $\phi=0.8433\approx \phi_J$.  In Fig.~\ref{g-x}(a) we show results for the larger strain rates $\dot\gamma\ge 5\times 10^{-7}$. Our results are from systems with $N=262144$ particles, except for the three largest strain rates, which use $N=65536$ particles. The solid lines are fits to Eq.~(\ref{exp2pbc}) using the form of Eq.~(\ref{exp2}) for $f(x)$.  We see clearly the nonmonotonic behavior as $x$ increases that requires the use of the two-exponential form of Eq.~(\ref{exp2}).

In Fig.~\ref{g-x}(b) we show results for $\dot\gamma\le 5\times 10^{-7}$.  For these smaller values of $\dot\gamma$, while $g(x)$ is  still seen to go negative, the minimum in $g(x)$ becomes very shallow and indeed seems to vanish at the smallest $\dot\gamma$.  As we will see below, the length $\ell$ has grown large, almost to the  size $L/2$, while the coefficient ratio $B/A$ is decreasing.  It is thus numerically unstable to try to fit to Eq.~(\ref{exp2}) and determine $\ell$.  Nevertheless we can still hope to determine $\xi$ from the initial decay of $g(x)$. For this we empirically fit $g(x)$ to the simpler form given by Eq.~(\ref{exp1}) below,  
\begin{equation}
f(x)=A\mathrm{e}^{-x/\xi}-\bar B.
\label{exp1}
\end{equation}
Such fits give the solid lines in Fig.~\ref{g-x}(b).

\begin{figure}
\centering
\includegraphics[width=3.3in]{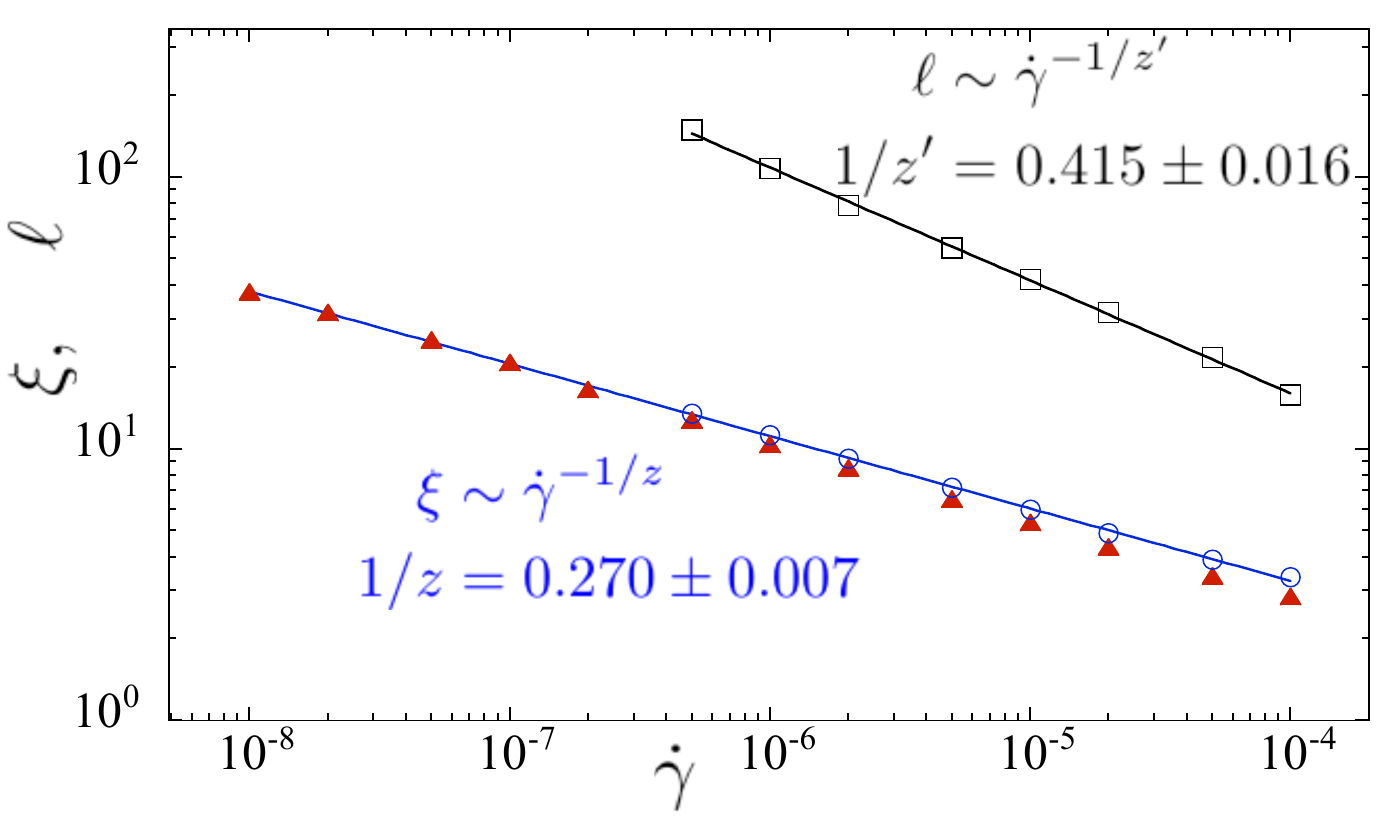}
\caption{Diverging length scales $\xi$ and $\ell$ vs strain rate $\dot\gamma$ at $\phi=0.8433\approx \phi_J$.  Open symbols  result from fits to the form of Eq.~(\ref{exp2}), while closed triangles for $\xi$ result from fits to the form of Eq.~(\ref{exp1}). 
}
\label{lengths} 
\end{figure}

In Fig.~\ref{lengths} we plot the resulting values of $\xi$ and $\ell$, as obtained from the fits described above.  For $\dot\gamma\ge5\times 10^{-7}$ we use the two-exponential form of Eq.~(\ref{exp2}) to determine both $\xi$ and $\ell$.  For $\dot\gamma < 5\times 10^{-7}$ we use the simpler form of Eq.~(\ref{exp1}) to determine $\xi$.  For comparison, and to indicate how well we might expect Eq.~(\ref{exp1}) to do, we also show results for $\xi$ obtained at larger $\dot\gamma$ by fitting to Eq.~(\ref{exp1}), but limiting the data used in the fit to $x<0.8x_\mathrm{min}$.  We see that the values of $\xi$ obtained from this simpler fit of Eq.~(\ref{exp1}) tend to be slightly smaller than those obtained from Eq.~(\ref{exp2}), but that the two approach each other as $\dot\gamma$ decreases.  This gives us confidence that the values of $\xi$ obtained at low $\dot\gamma$  via Eq.~(\ref{exp1}) are reasonable.

We see that for both $\xi$ and $\ell$, the data in Fig.~\ref{lengths} fall on a nice straight line, giving a power-law divergence for each length, $\xi\sim\dot\gamma^{-1/z}$ with $1/z=0.270\pm0.007$, and $\ell\sim\dot\gamma^{1/z^\prime}$ with $1/z^\prime=0.415\pm 0.016$.
Using $1/z\nu=0.26\pm0.02$ \cite{OT2} we then get a correlation length exponent of $\nu=(1/z)/(1/z\nu)=1.04\pm0.08$ for $\xi$, and $\nu^\prime=(1/z^\prime)/(1/z\nu)=1.60\pm 0.14$ for $\ell$.  The length $\ell$ thus diverges more rapidly than the length $\xi$, while $\xi$ diverges with an exponent consistent with $\nu=1$.  This is the main result of this work.    Note, by construction, $z^\prime\nu^\prime=z\nu$.

\section{Scaling Analysis}
\label{scale}

In this section we address the question of how
there can be two different diverging length scales $\xi$ and $\ell$, with different critical exponents.  We start with a usual scaling ansatz for the correlation function \cite{Chaikin},
\begin{equation}
g(x)=b^s \mathcal{G}(\delta\phi b^{1/\nu},\dot\gamma b^z, x b^{-1}, wb^{-\omega}).
\label{scaleg}
\end{equation}
As with Eq.~(\ref{scalexi}), $\nu$ is the correlation length exponent, $z$ the dynamic critical exponent, and $b$ an arbitrary length rescaling factor.  Since the separation $x$ is a length, it must scale in the combination $x/b$.  We also add to Eq.~(\ref{scaleg}) the leading correction-to-scaling variable $w$ \cite{OT2,VagbergOlssonTeitel,Berthier}.  Since, in the scaling sense, $w$ is an irrelevant  variable, its scaling exponent $-\omega$  must be negative, so that the scaling variable $wb^{-\omega}$ vanishes in the limit of large length scales, $b\to\infty$ \cite{Binder,Hasenbusch}.

If we now choose $b=\dot\gamma^{-1/z}$, then the above becomes
\begin{equation}
g(x)=\dot\gamma^{-s/z}\mathcal{G}\left(\frac{\delta\phi}{\dot\gamma^{1/z\nu}}, 1, x\dot\gamma^{1/z}, w\dot\gamma^{\omega/z}\right)
\label{scaleg2}
\end{equation}
If we assumed that the irrelevant variable $w$ can be ignored (i.e., set $w\to 0$), then we would  conclude that at $\delta\phi=0$, i.e., at $\phi_J$, 
the correlation $g(x)$ depends  on distance $x$ only through the term $x\dot\gamma^{1/z}$.  This would thus define the correlation length  as 
\begin{equation}
\xi\sim\dot\gamma^{-1/z}.  
\end{equation}
For the more general case of $\delta\phi\ne 0$, $\xi$ would scale as in Eq.~(\ref{xi2}).  This approach gives only a single diverging length scale $\xi$.

However, in the previous section we have found empirically that there are two diverging length scales, $\xi$ and $\ell$.  We can extract such a second diverging length scale from Eq.~(\ref{scaleg2}) if, instead of assuming $w$ can be ignored, we assume  that $w$ is a {\em dangerous irrelevant variable} \cite{Chaikin}, and that the scaling function $\mathcal{G}(\rho,1,u,v)$ contains a term proportional to $uv$.  In this case, when $\delta\phi=0$,  $g(x)$ will depend on distance $x$ through the two terms $u=x\dot\gamma^{1/z}$ and $uv=xw\dot\gamma^{(1+\omega)/z}$. A new diverging length scale 
\begin{equation}
\ell\sim \dot\gamma^{-(1+\omega)/z}/w
\label{ellw}
\end{equation}
thus appears.  Since $\omega$ must be positive, $\ell$ diverges more rapidly as $\dot\gamma\to 0$ than does $\xi$.  Since the irrelevant variable $w$ is presumed to be small, $\ell$ is large.  Both these conclusions are in accord with our findings in the previous section.


For $\delta\phi\ne 0$, $\ell$  should become independent  of $\dot\gamma$ as $\dot\gamma\to 0$.  Since the scaling function $\mathcal{G}$ depends on the packing via the variable $\delta\phi/\dot\gamma^{1/z\nu}$,  we conclude that, as $\dot\gamma\to 0$, 
$\ell\sim |\delta\phi|^{-(1+\omega)\nu}$.  Comparing with our notation of the previous section we thus have at $\phi=\phi_J$, 
\begin{equation}
\ell\sim\dot\gamma^{-1/z^\prime}\text{ with }z^\prime=z/(1+\omega)
\end{equation}
while for $\phi\ne\phi_J$ as $\dot\gamma\to 0$,
\begin{equation}
\ell\sim |\delta\phi|^{-\nu^\prime}\text{ with }\nu^\prime=(1+\omega)\nu.
\end{equation}
Using the values of $z$ and $z^\prime$ obtained in Fig.~\ref{lengths}, we get $\omega= 0.54\pm 0.03$.

\begin{figure}
\centering
\includegraphics[width=3.3in]{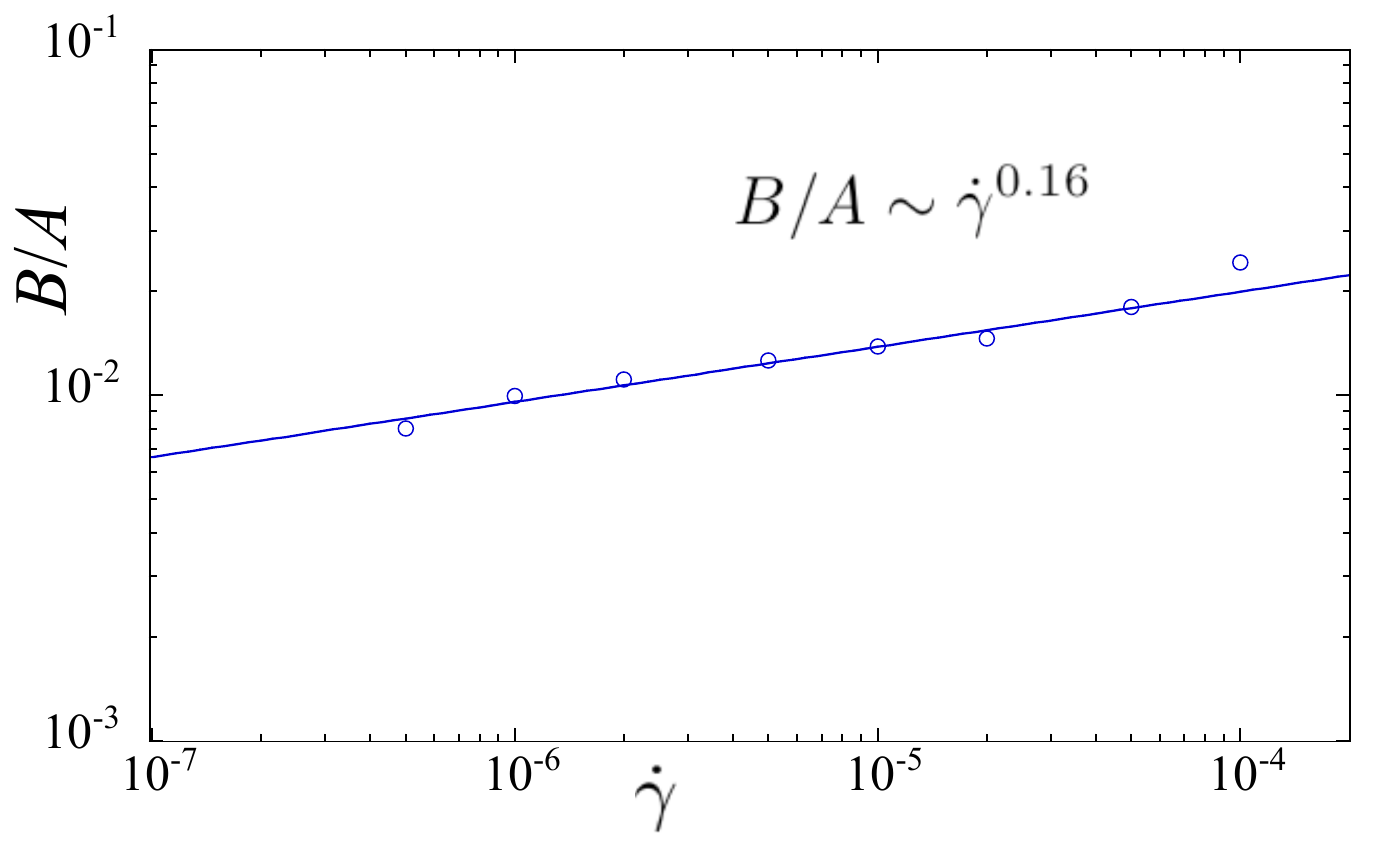}
\caption{Amplitude ratio $B/A$ of Eq.~(\ref{exp2}) vs $\dot\gamma$ at $\phi=0.8433\approx \phi_J$.  The straight line indicates a powerlaw behavior with exponent $0.16\pm0.04$.
}
\label{BA} 
\end{figure}

The above arguments yield several testable predictions.  
Assuming the scaling function is reasonably described by the empirical form of Eq.~(\ref{exp2}), then at $\delta\phi=0$ (i.e., $\phi=\phi_J$) the coefficient ratio $B/A$  can only depend on the scaling variable $w\dot\gamma^{\omega/z}$.  The simplest assumption is  that $B/A$ varies linearly in this variable.  In this case we expect $B/A\sim \dot\gamma^{\omega/z}$, with $\omega/z=(1/z^\prime)-(1/z)=0.145\pm 0.014$.
In Fig.~\ref{BA} we plot $B/A$ vs $\dot\gamma$ at $\phi=0.8433\approx\phi_J$.  We use only results for $\dot\gamma\ge 5\times 10^{-7}$, where we can fit $g(x)$ to the form of Eq.~(\ref{exp2}), and thus accurately determine the coefficient $B$ of the second exponential term that arises from the correction to scaling variable $w$.  We see a fair power-law behavior, $B/A\sim\dot\gamma^{0.16}$, with an exponent $0.16\pm 0.04$ in good agreement with $\omega/z= 0.145\pm 0.014$.

Thus, as $\dot\gamma\to 0$, we see that while $\ell$ diverges faster than the length $\xi$,  the exponential term involving the decay length $\ell$ becomes negligible compared to the  exponential term that decays with $\xi$.  
This is consistent with the assumption that the term involving $\ell$ arises  from an irrelevant variable $w$.

The preceding discussion has focused on behavior at the jamming $\phi_J$.  Another testable prediction involves behavior at $\phi\ne\phi_J$.  The scaling equation (\ref{scaleg2}) predicts that for $\delta\phi\ne 0$, the quantities $\xi$, $\ell$, and $B/A$  will be modified by  scaling functions that depend only on the variables $\delta\phi/\dot\gamma^{1/z\nu}$ and $w\dot\gamma^{\omega/z}$,
\begin{equation}
\xi \,\dot\gamma^{1/z}=h_\xi\left(\frac{\delta\phi}{\dot\gamma^{1/z\nu}}, w\dot\gamma^{\omega/z}\right),
\label{exi}
\end{equation}
\begin{equation}
\ell\, \dot\gamma^{(1+\omega)/z}=  h_\ell\left(\frac{\delta\phi}{\dot\gamma^{1/z\nu}}, w\dot\gamma^{\omega/z}\right),
\label{eell}
\end{equation}
\begin{equation}
(B/A)\, \dot\gamma^{-\omega/z} = \bar h\left(\frac{\delta\phi}{\dot\gamma^{1/z\nu}}, w\dot\gamma^{\omega/z}\right).
\label{eBA}
\end{equation}
If we assume that the irrelevant variable $w$ is sufficiently small that it can be neglected in the above (i.e. $w\to 0$), then we expect that plotting the left-hand side of each of Eqs.~(\ref{exi})--(\ref{eBA}) vs $\delta\phi/\dot\gamma^{1/z\nu}$ will result in a collapse of the data to a common curve.

\begin{figure}  
\centering
\includegraphics[width=3.3in]{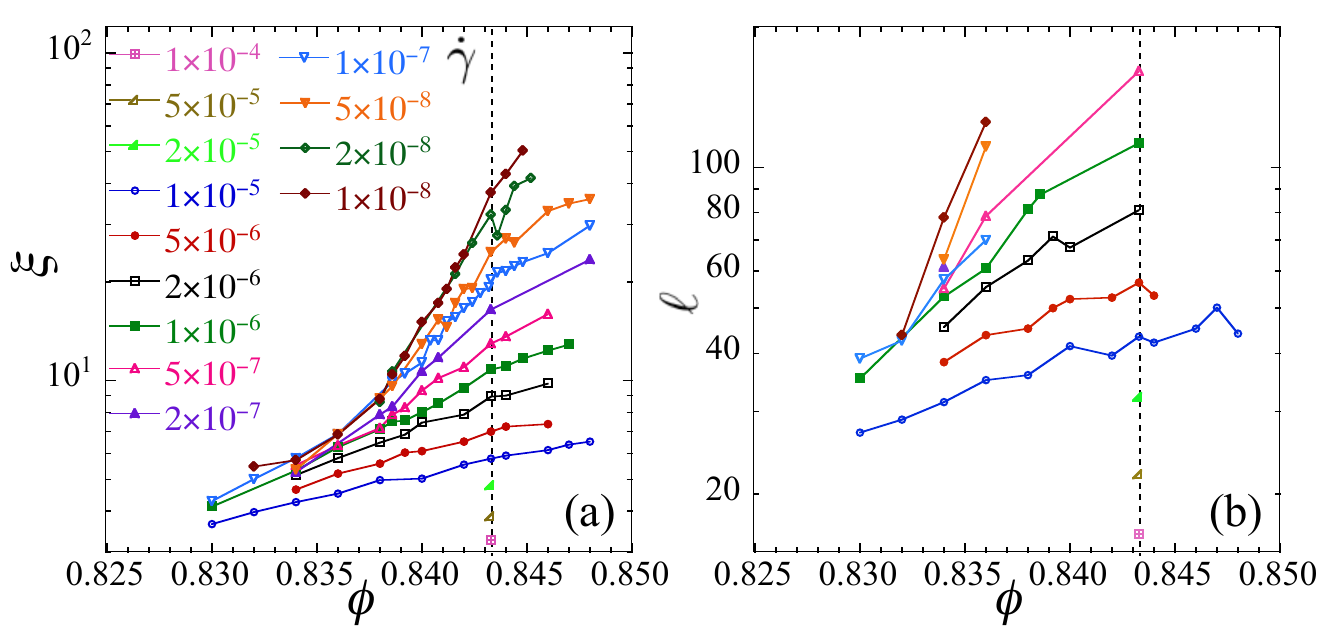}
\caption{Lengths (a) $\xi$ and (b) $\ell$ vs packing $\phi$ for different shear strain rates $\dot\gamma$.  The vertical dashed lines locate the jamming transition at $\phi_J\approx 0.8433$.  The symbols in (b) follow the same legend as in (a).
}
\label{xi-ell} 
\end{figure}

\begin{figure}
\centering
\includegraphics[width=3.3in]{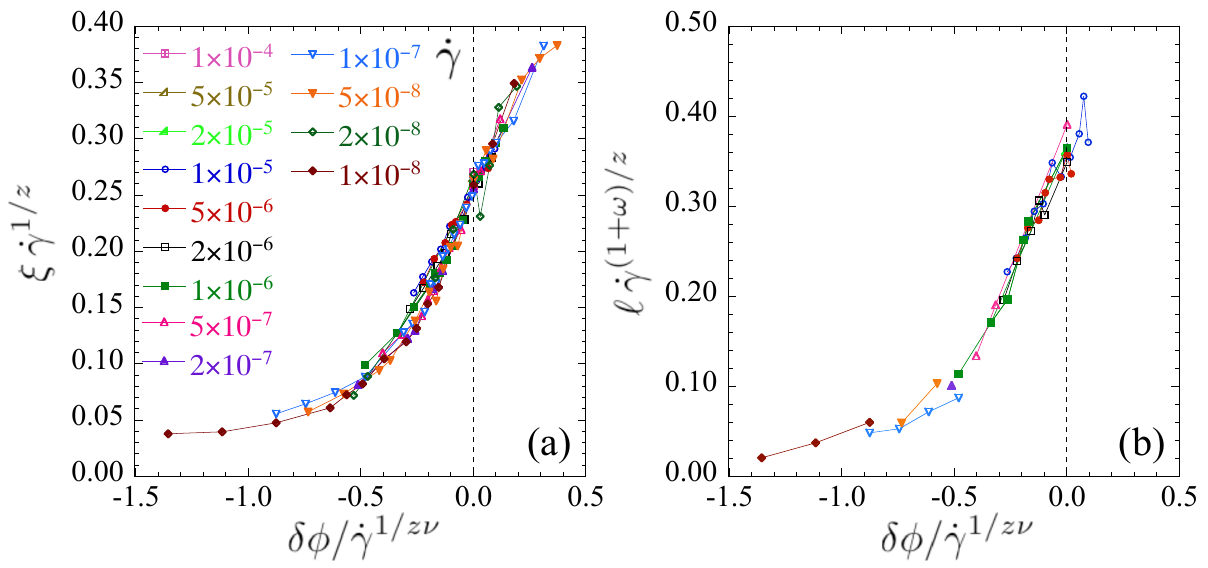}
\caption{Scaled lengths (a) $\xi\,\dot\gamma^{1/z}$ and (b) $\ell\,\dot\gamma^{(1+\omega)/z}$ vs scaled packing $\delta\phi/\dot\gamma^{1/z\nu}$, for different shear strain rates $\dot\gamma$.  The vertical dashed lines locate the jamming transition at $\delta\phi=0$ ($\phi=\phi_J$).  Plots are made using $\phi_J=0.8433$, $1/z=0.27$, $(1+\omega)/z=0.415$, and $1/z\nu=0.26$.  The symbols in (b) follow the same legend as in (a).
}
\label{xi-ell-scaled} 
\end{figure}

In Fig.~\ref{xi-ell}(a) we plot $\xi$ vs $\phi$ for different strain rates $\dot\gamma$.  The values of $\xi$ come from fits to either Eq.~(\ref{exp2}) or (\ref{exp1}), as needed.  In Fig.~\ref{xi-ell}(b) we plot $\ell$ vs $\phi$.  The values of $\ell$ come from fits to only Eq.~(\ref{exp2}). 
Just as  in Fig.~\ref{lengths} at $\phi_J$,  there  are considerably fewer data points for $\ell$ than for $\xi$ since fits to  Eq.~(\ref{exp2})   become unreliable as  $\ell$ gets large.  In Fig.~\ref{xi-ell-scaled} we show the corresponding scaled plots of $\xi\,\dot\gamma^{1/z}$ and $\ell\,\dot\gamma^{(1+\omega)/z}$ vs $\delta\phi/\dot\gamma^{1/z\nu}$.  As expected, we find a reasonable data collapse in both cases.  In Fig.~\ref{xi-ell-scaled}(a), where there are more data points, we see that the curves for different $\dot\gamma$ slightly increase, away from the $\dot\gamma\to 0$ limiting curve, as $\dot\gamma$ increases.  This is an indication that, for the larger $\dot\gamma$, the correction-to-scaling variable $w$ is not quite negligible.

\begin{figure}
\centering
\includegraphics[width=3.3in]{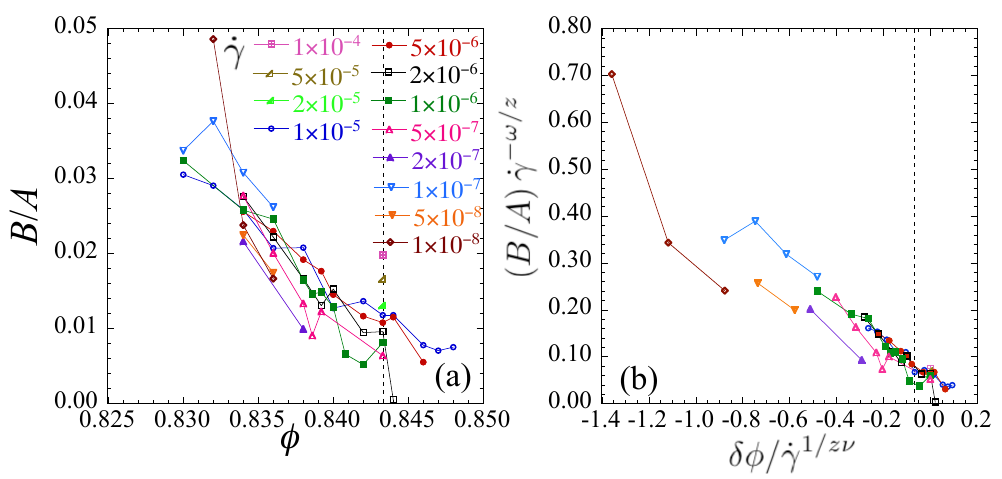}[t]
\caption{(a) Amplitude ratio $B/A$ vs packing $\phi$ for different shear strain rates $\dot\gamma$.  The vertical dashed line locates the jamming transition at $\phi_J\approx 0.8433$. (b) Scaled amplitude ratio $(B/A)\,\dot\gamma^{-\omega/z}$ vs scaled packing $\delta\phi/\dot\gamma^{1/z\nu}$, for different $\dot\gamma$.  The dashed line locates the jamming transition $\delta\phi=0$.  Plots are made using $\phi_J=0.8433$, $\omega/z=0.145$, and $1/z\nu=0.26$.  The symbols in (b) follow the same legend as in (a).
}
\label{BA-all} 
\end{figure}

In Fig.~\ref{BA-all}(a) we plot the coefficient ratio $B/A$ vs $\phi$ for different strain rates $\dot\gamma$.  The results for $A$ and $B$ used here come only from fits using Eq.~(\ref{exp2}).  Compared to our data for $\xi$ and $\ell$ in Fig.~\ref{xi-ell},  the data for $B/A$ are considerably noisier.  In Fig.~\ref{BA-all}(b) we show the corresponding scaled plot of $(B/A)\dot\gamma^{-\omega/z}$ vs $\delta\phi/\dot\gamma^{z\nu}$.   The collapse is similarly less satisfactory than the corresponding collapses for $\xi$ and $\ell$ in Fig.~\ref{xi-ell-scaled}.  For this collapse we have used the value of $\omega/z=(1/z^\prime)-(1/z)=0.145$, using  the values of $1/z$ and $1/z^\prime$ from the fits to $\xi$ and $\ell$ of Fig.~\ref{lengths}.  If we used instead $\omega/z=0.16$, from the fit to $B/A$ at $\phi_J$ of Fig.~\ref{BA}, then the collapse does not visibly improve.  

One reason that the scaled plot of $A/B$ fails to collapse nicely might be the effect of the correction-to-scaling variable $w$, as we have discussed above in connection with $\xi$.  But two other, probably more serious, reasons are the following.  (i) The fitting form of Eq.~(\ref{exp2}) is only an approximation to the true scaling function. Forcing the scaling function to fit to this form might skew results for the amplitudes $A$ and $B$ to a greater extent than for the length scales $\xi$ and $\ell$. (ii) Fitting to Eq.~(\ref{exp2}), and accurately determining the second exponential term, becomes difficult when $\ell$ is large.  Indeed, we see in Fig.~\ref{BA-all}(b) that it is the data at small rates $\dot\gamma <5\times 10^{-7}$ (the same range where we cannot determine $\ell$ when $\phi=\phi_J$) where the data depart most from a common curve; the data for $\dot\gamma\ge 5\times 10^{-7}$ collapses to a much better extent.  Thus we conclude that the behavior of $B/A$, while lacking the precision of other quantities, is consistent with our scaling analysis. 


There is one troubling aspect of our analysis above.  Here we have argued that the correction to scaling exponent found from $g(x)$ is $\omega/z= 0.145\pm 0.014$.  However, in our earlier critical scaling analysis of the pressure $p$ \cite{OT2}, that included corrections to scaling, we found $\omega/z= 0.29\pm0.03$.  Thus the $\omega$ found in this work is half the value found from our scaling analysis of pressure.  It could be that the leading correction-to-scaling variable that effects $g(x)$ is different from the one that effects $p$.  Or it could be that the scaling equation for  $p$, 
\begin{equation}
p=\dot\gamma^q h_p\left(\frac{\delta\phi}{\dot\gamma^{1/z\nu}}, w\dot\gamma^{\omega/z}\right),
\label{pscale}
\end{equation}
is such that when one expands the scaling function $h_p(\rho,v)$ about $v=0$, the leading term is proportional to $v^2$, rather than $v$.  This would cause the correction-to-scaling term in the scaling of $p$ to scale as $\dot\gamma^{2\omega/z}$, and  therefore reconcile that analysis with the present one. 

\section{Relation between $\xi$ and $p/\dot\gamma$}
\label{eta}

In  Sec.~\ref{newC}  we identified two diverging length scales, $\xi$ and $\ell$.  In  Sec.~\ref{scale} we presented a scaling analysis that indicated that, while $\ell$ diverges more rapidly than $\xi$ as the jamming transition is approached, $\ell$ arises from an ``irrelevant" (in the renormalization group sense) variable.  Thus we  expect  that it  is $\xi$ that is the correlation length that determines the singular behavior of global quantities at jamming.  Here we present further evidence to support this view.

\begin{figure}
\centering
\includegraphics[width=3.3in]{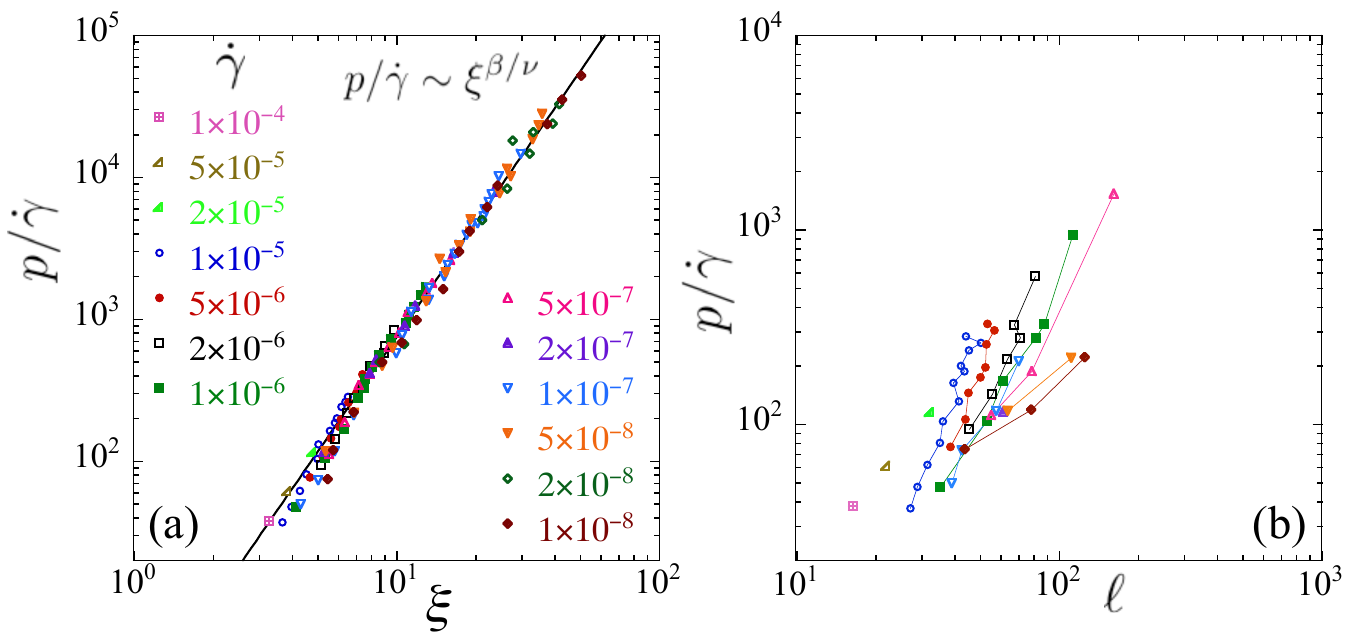}
\caption{(a) Plot of transport coefficient $p/\dot\gamma$ vs $\xi$ for different strain rates $\dot\gamma$ and different packings $\phi$.   The solid line is a power-law fit of the data to $p/\dot\gamma\sim\xi^{\beta/\nu}$ for $\xi >6$ and yields  $\beta/\nu=2.68\pm 0.08$.  (b) Plot of $p/\dot\gamma$ vs $\ell$ for different $\dot\gamma$ and $\phi$.  No simple relationship is revealed.  The symbols in (b) follow the same legend as in (a).
}
\label{eta-xi} 
\end{figure}

Consider the transport coefficient  $p/\dot\gamma$, which is the pressure analog of the shear viscosity.  As $\dot\gamma\to 0$, in the liquid-like phase below $\phi_J$, this transport coefficient diverges algebraically, $p/\dot\gamma \sim (\phi_J-\phi)^{-\beta}$, as the jamming transition is approached \cite{OT1,OT2,VagbergOlssonTeitel}.  This divergence is due to the diverging correlation length.
In Fig.~\ref{eta-xi}(a) we plot $p/\dot\gamma$ vs the length $\xi$, and in Fig.~\ref{eta-xi}(b) we plot $p/\dot\gamma$ vs the length $\ell$.  The data in these figures are at the same set of $(\phi,\dot\gamma)$ values as in Fig.~\ref{xi-ell}.
When plotting vs $\xi$, the data for $p/\dot\gamma$ give an excellent collapse to a common curve with a simple  power-law relation,  
\begin{equation}
p/\dot\gamma\sim\xi^{\beta/\nu}.
\label{eetaxi}
\end{equation}
Fitting the data for $\xi>6$ gives  the exponent $\beta/\nu=2.68\pm 0.08$.  However, there is no collapse when plotting vs $\ell$. This indicates that it is $\xi$ and not $\ell$ that controls the divergence of $p/\dot\gamma$, as one approaches the jamming transition.

The scaling behavior of $p/\dot\gamma$ follows from that of Eq.~(\ref{pscale}).  Assuming that, to lowest order, the correction-to-scaling variable $w$ may be ignored, we have,
\begin{equation}
p/\dot\gamma=\dot\gamma^{-\beta/z\nu}h_p\left(\frac{\delta\phi}{\dot\gamma^{1/z\nu}},0\right),
\label{etascale}
\end{equation}
where $\beta/z\nu = 1-q$.  For $\delta\phi=0$, we thus have $p/\dot\gamma\sim\dot\gamma^{-\beta/z\nu}$.  Since $\xi\sim\dot\gamma^{-1/z}$ when $\delta\phi=0$, it then follows that $p/\dot\gamma\sim \xi^{\beta/\nu}$ when $\phi=\phi_J$.  For $\phi<\phi_J$, we know that $p/\dot\gamma$ 
has a finite limit as $\dot\gamma\to 0$; it thus must be true that $h_p(u\to\infty,0)\sim |u|^{-\beta}$, so that the scaling Eq.~(\ref{etascale}) yields $p/\dot\gamma\sim |\delta\phi|^{-\beta}$ as $\dot\gamma\to 0$.  In this same limit we have $\xi\sim |\delta\phi|^{-\nu}$.  
Thus, as $\dot\gamma\to 0$ for $\phi<\phi_J$, we again recover $p/\dot\gamma\sim\xi^{\beta/\nu}$.
Using the values $\beta/\nu=2.68\pm 0.08$ from the fit in  Fig.~\ref{eta-xi}(a), and $\nu=1$ from our results of Sec.~\ref{newC}, we thus get $\beta = 2.68\pm 0.08$. This is in excellent agreement with our earlier results from a direct scaling analysis of the rheology \cite{OT2,OT3}.  

Note that, unlike the scaling analysis of Refs.~\cite{OT2,OT3}, the analysis of Fig.~\ref{eta-xi}(a) allows the determination of $\beta/\nu$ without the need to know the value of $\phi_J$.  In an earlier work \cite{Olsson1}, one of us established that, for $\phi<\phi_J$, $p/\dot\gamma$ scales the same as the relaxation time $\tau$ that describes the decay of a sheared configuration to zero energy, once the shearing has been turned off.  Since it is known that the system is isostatic at jamming, with the average contact number per particle $z_\mathrm{iso}=2d=4$ in two dimensions, plotting $\tau$ vs $\delta z=z_\mathrm{iso}-z$ yields a power-law behavior $\tau\sim \delta z^{-\beta/u_z}$. Here $z$ is the average contact number of the energy relaxed state. Thus one can determine the exponent $\beta/u_z$, again without having to know the value of $\phi_J$. Here $u_z$ is the exponent that determines how the contact number varies as the packing $\phi$ decreases below $\phi_J$, $\delta z\sim (\phi_J-\phi)^{u_z}$. The analysis in Ref.~\cite{Olsson1} gave $\beta/u_z=2.69\pm 0.03$, while that in Ref.~\cite{Lerner} gave $\beta/u_z=1/0.38=2.63$.  Comparing to the value $\beta/\nu=2.68\pm 0.08$ found here yields the conclusion $\nu\approx u_z$, and so $\xi\sim 1/\delta z$, in agreement with the earlier results of Ref.~\cite{Heussinger1}.  Taking $\nu=1$ from our analysis of Sec.~\ref{newC}, we also conclude that $u_z=1$, again recovering  earlier results of Ref.~\cite{Heussinger1}.

While, from the above arguments, we expect Eq.~(\ref{eetaxi}) to hold both exactly at $\phi_J$, and for $\dot\gamma\to 0$ below $\phi_J$, it is surprising to find in Fig.~\ref{eta-xi}(a) that  this relation seems to hold more generally, for any $\phi$ and $\dot\gamma$.
Comparing Eqs.~(\ref{exi}) with (\ref{etascale}), we see that for this relation to hold in general, it is  necessary that the respective scaling functions for $p/\dot\gamma$ and for $\xi$ obey the relation,
\begin{equation}
h_p(u,0)=[h_\xi(u,0)]^{\beta/\nu}.
\label{hphxi}
\end{equation}
In general,  a scaling approach does not assume any knowledge about the details of the scaling function, or the relation between scaling functions of different quantities as in Eq.~(\ref{hphxi}), except for  behaviors in different asymptotic limits.  However the more general result of Eq.~(\ref{eetaxi}) can be shown to follow from an effective density approximation that we have introduced previously \cite{OT3}, and which we have found to describe well the rheology of the system, provided one does not go too far above $\phi_J$.  We have found that behavior at a given $\phi$ and $\dot\gamma$ is well described by considering the system to be in the hard-core $\dot\gamma\to 0$ limit, but at an effective packing given by $\phi_\mathrm{eff}(\phi,\dot\gamma)=\phi-cE^{1/2y}$, where $E(\phi,\dot\gamma)$ is the elastic energy of the system at the given packing and strain rate, $c=1.54$ and $y=1.09$.  

\begin{figure}
\centering
\includegraphics[width=3.3in]{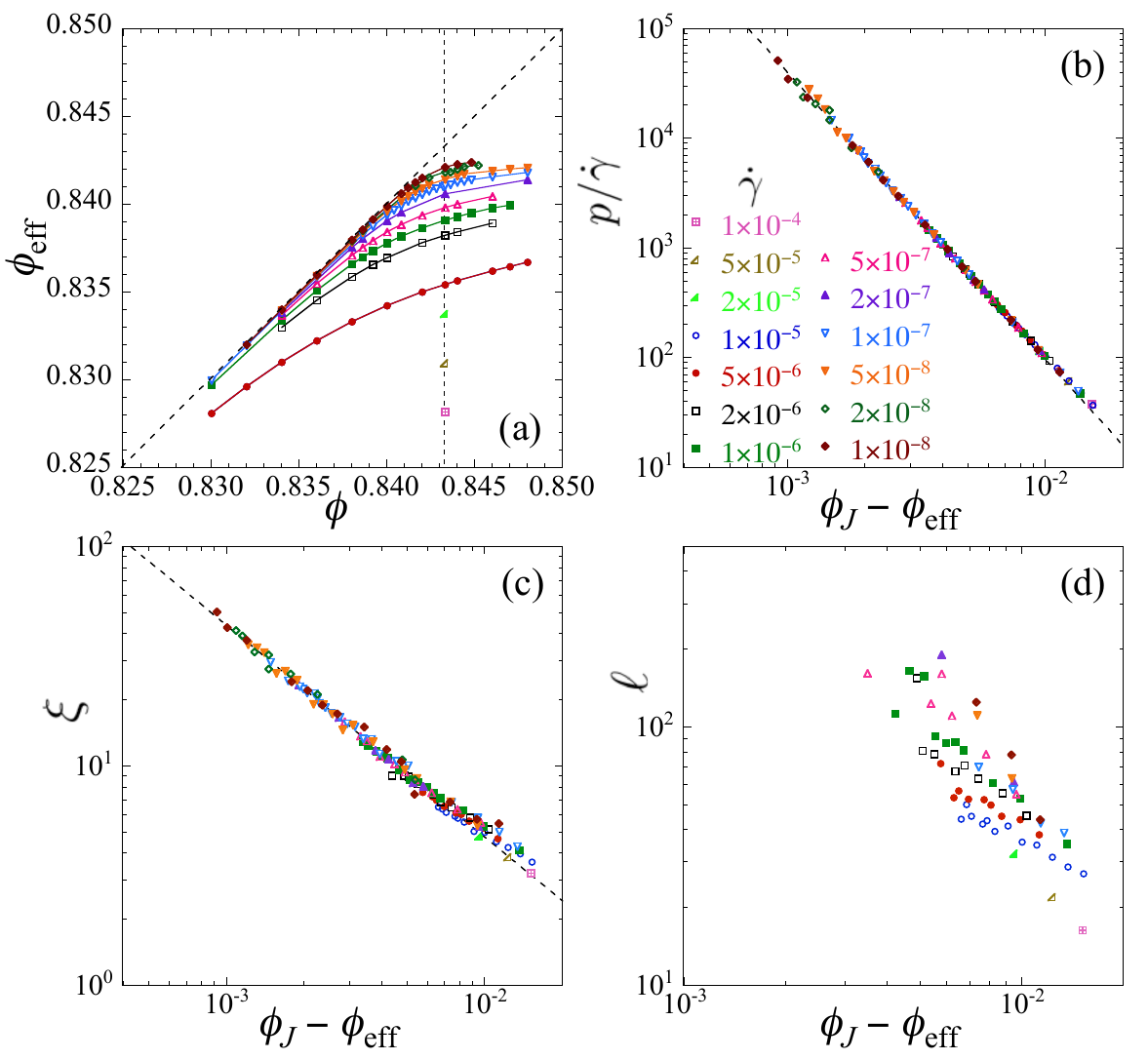}
\caption{(a) Effective packing $\phi_\mathrm{eff}(\phi,\dot\gamma)=\phi-cE^{1/2y}$ vs $\phi$ for different strain rates $\dot\gamma$; the diagonal dashed line denotes $\phi_\mathrm{eff}=\phi$, while the vertical dashed line locates $\phi_J\approx 0.8433$.  (b) $p/\dot\gamma$ vs $\phi_J-\phi_\mathrm{eff}$, (c) $\xi$ vs $\phi_J-\phi_\mathrm{eff}$, and (d) $\ell$ vs $\phi_J-\phi_\mathrm{eff}$, for different strain rates $\dot\gamma$.  The value $\phi_J=0.8433$ is used in making these plots.  In (b) and (c) the dashed line is a powerlaw fit to the data for $\phi_J-\phi_\mathrm{eff}\le 0.005$, and gives the exponent $\beta=2.62$ for $p/\dot\gamma$ and $\nu=0.98$ for $\xi$.  The symbols in (a), (c), and (d) follow the same legend as in (b).  
}
\label{phi-eff} 
\end{figure}

In Fig.~\ref{phi-eff}(a) we plot $\phi_\mathrm{eff}$ vs $\phi$ for different $\dot\gamma$.  For $\phi<\phi_J$, the smaller the $\dot\gamma$, the closer one is to the hard-core limit, and the smaller is the deviation of $\phi_\mathrm{eff}$ from $\phi$.  In Figs.~\ref{phi-eff}(b), \ref{phi-eff}(c), and \ref{phi-eff}(d) we plot $p/\dot\gamma$, $\xi$, and $\ell$, respectively, vs $\phi_J-\phi_\mathrm{eff}$, using $\phi_J=0.8433$.  We see that the data for $p/\dot\gamma$ and $\xi$ collapse to a nice power-law scaling,
\begin{align}
p/\dot\gamma&\sim (\phi_J-\phi_\mathrm{eff})^{-\beta}\\
\xi &\sim (\phi_J-\phi_\mathrm{eff})^{-\nu},
\end{align}
from which   the general result of Eq.~(\ref{eetaxi}) follows for any $\phi$ and $\dot\gamma$.   For $\ell$ we find no such nice collapse.

{\color{black}We  have empirically found that our effective density approximation describes well the leading critical singularity, but does not describe well effects due to corrections to scaling.  This is the reason we have considered $p/\dot\gamma$ here instead of the related shear viscosity $\sigma/\dot\gamma$.  In \cite{OT2} we demonstrated that corrections to scaling are considerably smaller for $p/\dot\gamma$ than they are for $\sigma/\dot\gamma$.}
That the data in Figs.~\ref{eta-xi}(a), \ref{phi-eff}(b) and \ref{phi-eff}(c) are simple power-law relations is a signature that, for these quantities, the corrections to scaling are generally small for our range of data.  However, from Eq.~(\ref{ellw}) we see that  the length $\ell$ would  diverge in the limit of $w\to 0$, when corrections to scaling vanish.  Thus, unlike $p/\dot\gamma$ and $\xi$ which have well defined limiting behaviors as $w\to 0$, the length $\ell$ requires the corrections to scaling to be finite in order for $\ell$ to be finite.  We believe this is the reason that $\ell$ in Fig.~\ref{phi-eff}(d) shows no nice collapse when plotted vs $\phi_\mathrm{eff}$.

\section{Rotation and divergence of the velocity field}
\label{vorticity}

\begin{figure}
  \includegraphics[width=8cm]{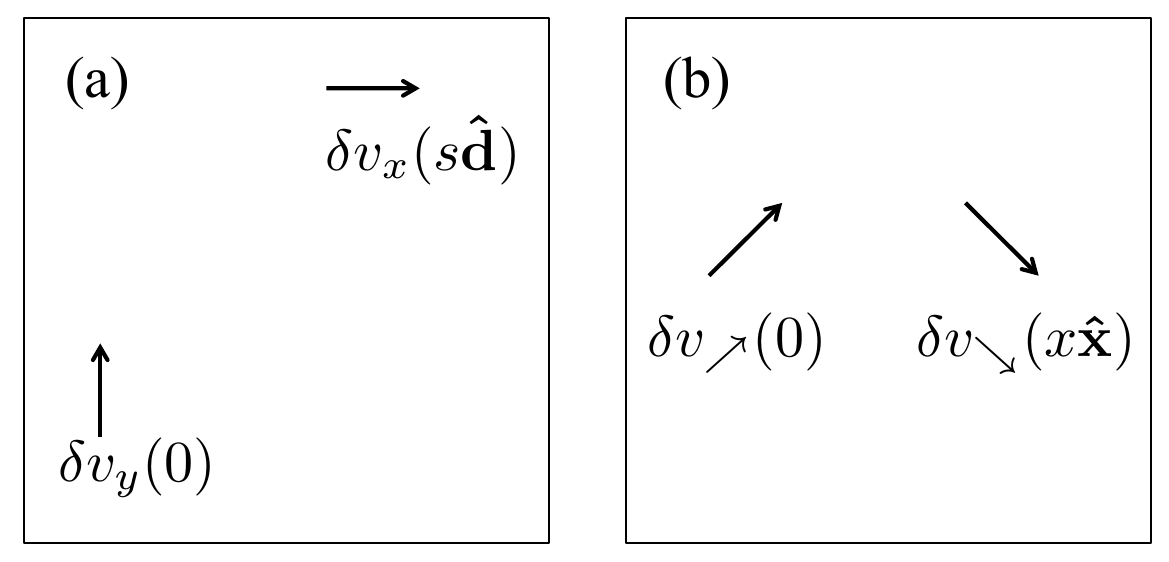}
  \caption{Schematic of the terms that contribute to the velocity correlations of (a) Eq.~(\ref{eq:vxvy-xy}) and (b) Eq.~(\ref{vnevse}).}
  \label{vv-corr}
\end{figure}

In this section we provide a physical interpretation for the particular velocity correlation  $g(x)$ of Eq.~(\ref{gmix}), as well as a physical interpretation of the lengths $\xi$ and $\ell$.  
We first consider the \emph{mixed} correlation of different velocity components,
$\langle \delta v_y(0) \delta v_x(\mathbf{r})\rangle$, where again $\delta\mathbf{v}=\mathbf{v}-\dot\gamma y\mathbf{\hat x}$ is the nonaffine part of the particle velocity field, i.e., the fluctuation of the  velocity  away from a uniform shear flow. For $\mathbf{r}=x\mathbf{\hat x}$ along the flow direction, this correlation vanishes by symmetry, 
 $ \langle \delta v_y(0) \delta v_x(x\mathbf{\hat x})\rangle=0$.
However, this is not the case for $\mathbf{r}$ along the system diagonal. With $\mathbf{\hat d}= (\mathbf{\hat x} + \mathbf{\hat 
y})/\sqrt 2$ the unit vector in the diagonal direction, the correlation of the components, 
\begin{equation}
  \langle \delta v_y(0) \delta v_x(s\mathbf{\hat d})\rangle,
  \label{eq:vxvy-xy}
\end{equation}
as shown in Fig.~\ref{vv-corr}(a),
is in general nonvanishing. The geometry of this correlation suggests that it is a measure of the rotation of  fluctuations of the velocity field   \cite{Olsson2}.


In our shearing geometry, with periodic boundary conditions in the $\mathbf{\hat x}$ direction, and Lees-Edwards boundary conditions in the $\mathbf{\hat y}$ direction, it is easiest to measure correlations along the $\mathbf{\hat x}$ direction, rather than along $\mathbf{\hat d}$.  We  therefore define  correlations similar to Eq.~(\ref{eq:vxvy-xy}) as follows.  With $\mathbf{\hat d}$ as defined above, and the orthogonal unit vector $\mathbf{\hat d}_\perp=(\mathbf{\hat x} - \mathbf{\hat y})/\sqrt 2$, we define
\begin{align}
\delta v_\nearrow &=\delta \mathbf{v}\cdot\mathbf{\hat d}=(\delta v_x+\delta v_y)/\sqrt{2}, \label{vne}\\
\delta v_\searrow &=\delta \mathbf{v}\cdot\mathbf{\hat d}_\perp=(\delta v_x-\delta v_y)/\sqrt{2}\label{vse}.
\end{align}
We can then consider the correlations,
\begin{equation}
\langle \delta v_\nearrow(0) \delta v_\searrow (x\mathbf{\hat x})\rangle \,\,\text{and}\,\, \langle \delta v_\searrow(0)\delta v_\nearrow(x\mathbf{\hat x})\rangle.
\label{vnevse}
\end{equation}
We find numerically that these  two correlations are equal.
The first of these correlations,  shown in Fig.~\ref{vv-corr}(b),  is  related to that of Eq.~(\ref{eq:vxvy-xy}) by making a  clockwise rotation by $45^\circ$.  
Substituting Eqs.~(\ref{vne}) and (\ref{vse}) into Eq.~(\ref{vnevse}), and comparing with Eq.~(\ref{gmix}), we then find that
\begin{equation}
g(x)=\dfrac{\langle \delta v_\nearrow(0) \delta v_\searrow (x\mathbf{\hat x})\rangle + \langle \delta v_\searrow(0)\delta v_\nearrow(x\mathbf{\hat x})\rangle}{\langle|\delta\mathbf{v}|^2\rangle/2}.
\end{equation}
Our alternative velocity correlation $g(x)$ is thus a measure of the rotation of the velocity-field fluctuations.

This observation also suggests an interpretation for the  the second exponential term of Eq.~(\ref{exp2}), $-B \mathrm{e}^{-x/\ell}$, which gives a negative contribution to $g(x)$. That this term is negative implies that there is a contribution to the correlation $\langle\delta v_\nearrow(0)
  \delta v_\nwarrow(x\mathbf{\hat x})\rangle$ that is positive; such a term is related to the divergence of the velocity-field fluctuations.  
Our interpretation is thus that $\xi$ measures the 
size of  fluctuations in the rotation of the velocity field, whereas $\ell$ measures the
size of fluctuations in the divergence of the velocity field.
We now proceed to demonstrate this by a direct calculation.

In principle one would like to directly compute the correlations of the  rotation, $\mathbf{\hat z}\cdot \nabla\times\delta\mathbf{v}$, and divergence, $\nabla\cdot\delta \mathbf{v}$, of velocity fluctuations.  However that would require both a discretization of the velocity field to a grid, and then also the discretization of the velocity derivates.  Instead, we take a different approach.  We consider a circular window $\mathcal{C}_r$ of radius $r$ centered about a  point $\mathbf{r}_0$.  We then compute the integrals of the rotation and the divergence of $\delta\mathbf{v}$ over the area of $\mathcal{C}_r$
\begin{equation}
\Omega_r =\int_{\mathcal{C}_r}d^2r\, \mathbf{\hat z}\cdot \nabla\times\delta\mathbf{v}=\oint_{\Gamma_r}dl\,\mathbf{\hat t}\cdot \delta\mathbf{v}
\end{equation}
and
\begin{equation}
D_r = \int_{\mathcal{C}_r}d^2r\,\nabla\cdot\delta \mathbf{v}=\oint_{\Gamma_r}dl\,\mathbf{\hat n}\cdot\delta\mathbf{v},
\end{equation}
where $\Gamma_r$ is the circumference of $\mathcal{C}_r$, and $\mathbf{\hat t}$ and $\mathbf{\hat n}$ are the unit tangent and unit normal vectors to $\Gamma_r$.  Since the velocity field is only defined discretely at the sites of individual particles, we approximate the above integrals by
\begin{equation}
\bar\Omega_r= \sum_{i\in \Delta\Gamma_r}\mathbf{\hat t}\cdot\delta\mathbf{v}_i,\quad
\bar D_r = \sum_{i\in\Delta\Gamma_r}\mathbf{\hat n}\cdot\delta\mathbf{v}_i,
\end{equation}
where the sums are over all particles $i$ that lie within an annulus $\Delta\Gamma_r$ centered at radius $r$ and of thickness $\Delta r=1$.

Since $\langle\delta \mathbf{v}\rangle =0$, it necessarily follows that $\langle \bar\Omega_r\rangle = \langle \bar D_r\rangle=0$.  However we find that,
for sufficiently large $r$,  the mean-square fluctuations $\langle\bar\Omega_r^2\rangle$ and $\langle\bar D_r^2\rangle$ scale $\sim r$.  This can be rationalized if we think of $\delta\mathbf{v}$ as being a local  quantity that fluctuates independently on sufficiently large length scales; then both $\langle\bar\Omega_r^2\rangle$ and $\langle\bar D_r^2\rangle$ will scale proportional to the area of $\Delta\Gamma_r$.  
If we define $N_r$ as the number of particles within the annulus $\Delta\Gamma_r$, then similarly $\langle N_r\rangle\sim r$.  
The quantities $\langle\bar\Omega_r^2/N_r\rangle$ and $\langle\bar D_r^2/N_r\rangle$ then represent the mean-square fluctuations of the normal and tangential components of the velocity per particle within the annulus.  If velocity fluctuations were uncorrelated from particle to particle, and if velocity fluctuations were independent of orientation, then these quantities would both  be equal to $\langle |\delta\mathbf{v}|^2\rangle/2$.
We therefore define,
\begin{align}
C_\mathrm{rot}(r)&=\dfrac{1}{\langle |\delta\mathbf{v}|^2\rangle/2}\left\langle\dfrac{\bar\Omega_r^2}{N_r}\right\rangle,\\[12 pt]
C_\mathrm{div}(r)&=\dfrac{1}{\langle |\delta\mathbf{v}|^2\rangle/2}\left\langle\dfrac{\bar D_r^2}{N_r}\right\rangle,
\end{align}
where $\langle\dots\rangle$ denotes an average over both the center position  $\mathbf{r}_0$ of the circular window $\mathcal{C}_r$ within a given configuration, as well as over different configurations within the sheared steady state.    The difference of $C_\mathrm{rot}(r)$ and $C_\mathrm{div}(r)$ from unity is then a measure of the effect of velocity correlations.

\begin{figure}
  \includegraphics[width=8cm]{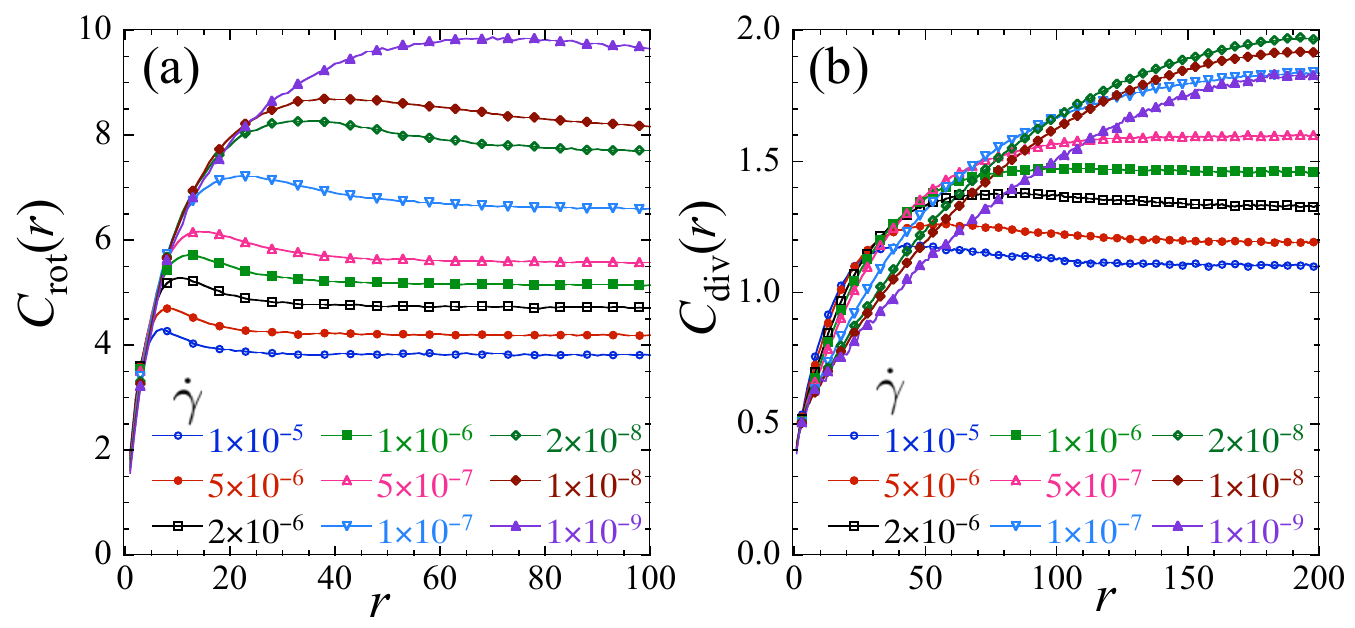}
  \caption{(a) $C_\mathrm{rot}(r)$ and (b) $C_\mathrm{div}(r)$ vs $r$ for different strain rates $\dot\gamma$ at fixed $\phi=0.8433\approx \phi_J$.  For clarity, symbols are only shown  on every fifth data point.
  }
  \label{Crot-div}
\end{figure}

\begin{figure}
  \includegraphics[width=8cm]{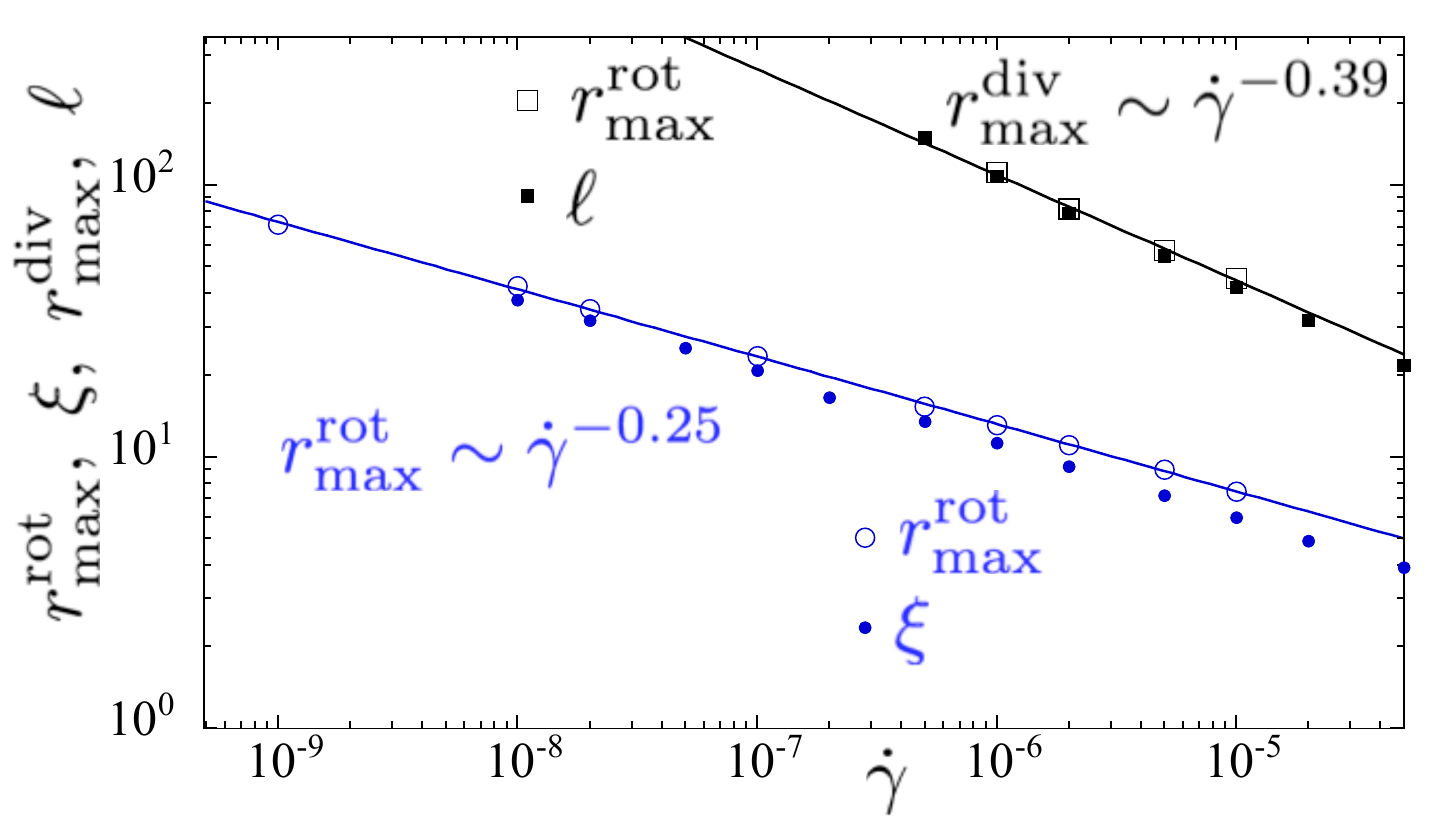}
  \caption{Open symbols: Location of the maximum $r_\mathrm{max}^\mathrm{rot}$ of $C_\mathrm{rot}(r)$, and location of the maximum $r_\mathrm{max}^\mathrm{div}$ of $C_\mathrm{div}(r)$, vs strain rate $\dot\gamma$ at fixed $\phi=0.8433\approx \phi_J$.  The solid lines are fits to a powerlaw with exponents $0.25$ and $0.39$, respectively.  Solid symbols:  length scales $\xi$ and $\ell$ as determined from the velocity correlation $g(x)$ in Sec.~\ref{newC}.
  }
  \label{x-rot-div-max}
\end{figure}

In Fig.~\ref{Crot-div}(a) we plot $C_\mathrm{rot}(r)$ vs $r$ for different strain rates $\dot\gamma$ at the fixed $\phi=0.8433\approx\phi_J$.  We see that as $r$ increases, $C_\mathrm{rot}(r)$ increases until it reaches a maximum, and then decreases a bit and plateaus to a constant.  The location of the maximum at $r_\mathrm{max}^\mathrm{rot}$  sets the length scale on which fluctuations saturate.  We see that $r_\mathrm{max}^\mathrm{rot}$ increases as $\dot\gamma$ decreases.  In Fig.~\ref{Crot-div}(b) we  plot $C_\mathrm{div}(r)$.  Here we see a similar behavior.  As $r$ increases, $C_\mathrm{div}(r)$ increases, reaches a maximum, and then plateaus to a constant.  The location of the maximum at $r_\mathrm{max}^\mathrm{div}$ increases as $\dot\gamma$ decreases.  However for the smallest $\dot\gamma \le5\times10^{-7}$, the curves continue to increase with $r$ and no maximum can be determined; this is because the relevant length scale has become too big compared to the finite length $L$ of our system.

In Fig.~\ref{x-rot-div-max} we plot the locations of these maxima, $r_\mathrm{max}^\mathrm{rot}$ and $r_\mathrm{max}^\mathrm{div}$, vs strain rate $\dot\gamma$ at fixed $\phi=0.8433$; these are shown as the open symbols.  The solid lines show power-law fits to these data, giving an exponent 0.25 for $r_\mathrm{max}^\mathrm{rot}$ and 0.39 for $r_\mathrm{max}^\mathrm{div}$.  These values are in reasonable agreement with the exponents  $1/z=0.27$ and $1/z^\prime=0.415$ found for the velocity  length scales $\xi$ and $\ell$ in Fig.~\ref{lengths} of Sec.~\ref{newC}.  To highlight this point, we also show as solid symbols in Fig.~\ref{x-rot-div-max} the data for $\xi$ and $\ell$ from Fig.~\ref{lengths}.  We clearly see that $r_\mathrm{max}^\mathrm{rot}\approx\xi$ and $r_\mathrm{max}^\mathrm{div}\approx\ell$.  

We thus conclude from this analysis that  the  correlations lengths $\xi$ and $\ell$, obtained from the velocity correlation $g(x)$, have the following interpretation.  The length $\xi$ measures the characteristic length scale of fluctuations in the rotation of the particle velocity field $\delta\mathbf{v}$, while $\ell$ measures the characteristic length scale of fluctuations in the divergence of $\delta\mathbf{v}$.

\begin{figure}
  \includegraphics[width=4.5cm]{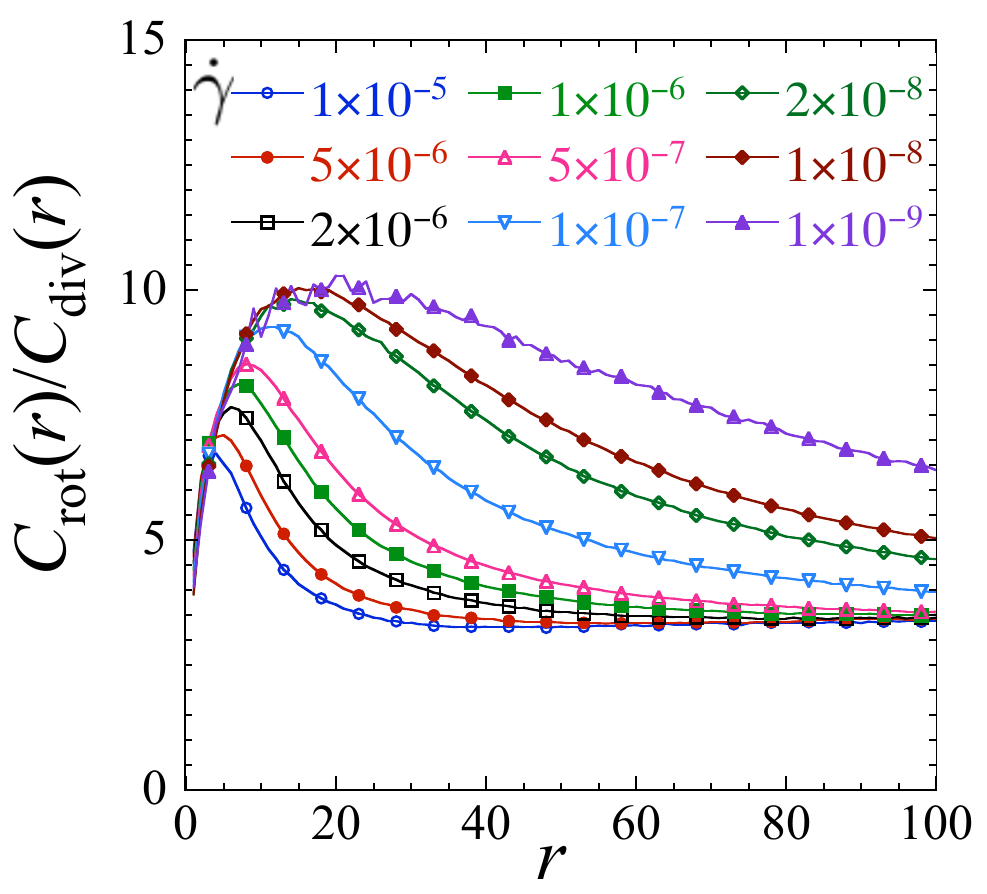}
  \caption{Ratio $C_\mathrm{rot}(r)/C_\mathrm{div}(r)$ vs $r$ for different strain rates $\dot\gamma$ at fixed $\phi=0.8433\approx \phi_J$.  Symbols are only shown on every fifth data point.
  }
  \label{CrCd}
\end{figure}

Note one additional point.  For $C_\mathrm{rot}(r)$ the curves for different $\dot\gamma$ appear to be approaching a common limiting curve in the hard-core limit, as $\dot\gamma \to 0$.  We find that this curve is logarithmic.  The  $\dot\gamma\to 0$ behavior of $C_\mathrm{div}(r)$ is less clear.  If we look at the ratio of $C_\mathrm{rot}(r)/C_\mathrm{div}(r)$, shown in Fig.~\ref{CrCd}, then we see that this ratio reaches a maximum at an $r_\mathrm{max}$ that increases with decreasing $\dot\gamma$, but that the large $r$ limit appears to be constant for all $\dot\gamma$, $\lim_{r\to\infty} C_\mathrm{rot}(r)/C_\mathrm{div}(r)\approx 3.5$.

\section{Summary}
\label{sum}

To summarize our results, we have considered the length scales that characterize the fluctuations of the particle velocity field $\delta\mathbf{v}$, in a simple-sheared system of athermal, bidisperse, frictionless circular disks in two dimensions, sheared at a constant rate $\dot\gamma$.  We have shown that our earlier analysis \cite{OT1} of the transverse velocity correlation function gave an erroneous value for the correlation length exponent $\nu$, because of a failure to appreciate the effects of multiple length scales.  We have then introduced an alternative velocity correlation function $g(x)$, related to the rotation of the velocity fluctuations, and find that it is reasonably characterized in terms of two different length scales $\xi$ and $\ell$.  By considering behavior for varying $\dot\gamma$ at the jamming $\phi_J$, we find numerically that these two lengths  diverge with different critical exponents, with $\ell$ diverging more rapidly than $\xi$ as the jamming point is approached.  

We then provide an analysis of $g(x)$ in terms of a critical scaling ansatz.  In terms of this scaling ansatz, $\ell$ is seen to arise from the effects of a dangerous irrelevant, correction-to-scaling, variable.  This scaling analysis provides  self-consistent results for both the relative amplitudes $B/A$ of the competing terms involving the two length scales, as well as for behavior when $\phi\ne\phi_J$.  Although the length $\ell$ diverges more rapidly than $\xi$, we find that the term involving $\ell$ scales to zero as jamming is approached.  The identification of $\ell$ with the effects of an irrelevant variable then leads to the conclusion that it is $\xi$ which is properly identified as the correlation length that controls the critical behavior at jamming, and we find results consistent with an exponent $\nu=1$, in agreement with an earlier finite-size scaling analysis \cite{VOT}.  The length $\ell$ diverges with the exponent $(1+\omega)\nu$, with the correction-to-scaling exponent $\omega\approx 0.5$.  We discussed how this value of $\omega$ can be reconciled with our earlier determination of $\omega$ in Ref.~\cite{OT2}.

Supporting our conclusion that $\xi$ is the correlation length, we  have shown how the pressure analog of viscosity, $p/\dot\gamma$, scales as a simple powerlaw of $\xi$, $p/\dot\gamma\sim \xi^{\beta/\nu}$, over a wide range of strain rates $\dot\gamma$ and packing fractions $\phi$.  We show that this result follows from an effective density approximation, in which the behavior of the system at  packing $\phi$ and finite strain rate $\dot\gamma$ can be mapped onto a hard-core system (i.e., $\dot\gamma\to 0)$ at an effective packing $\phi_\mathrm{eff}(\phi,\dot\gamma)$.  But we find no such simple relation between $p/\dot\gamma$ and $\ell$, nor is the behavior of $\ell(\phi,\dot\gamma)$ well described by the effective density approximation.

Finally we have provided the physical significance of the two lengths $\xi$ and $\ell$.  By directly considering the fluctuations of the rotation of the velocity field, $\mathbf{\hat z}\cdot\nabla\times\delta\mathbf{v}$, and the divergence of the velocity field $\nabla\cdot\delta\mathbf{v}$, we show that $\xi$ sets the length scale on which fluctuations in the rotation of $\delta\mathbf{v}$ saturate, while $\ell$ sets the length scale on which fluctuations in the divergence of $\delta\mathbf{v}$ saturate.

\section*{Acknowledgements}

This work was supported in part by National Science Foundation Grant No. DMR-1809318. The simulations were performed on resources provided by the Swedish National Infrastructure for Computing (SNIC) at HPC2N.


\bibliographystyle{apsrev4-1}

\begin{thebibliography}{99}

\bibitem{LiuNagel}A. J. Liu and S. R. Nagel,  The jamming transition and the marginally jammed solid, Annu. Rev. Condens. Matter Phys. {\bf 1}, 347 (2010).

\bibitem{OHern}C. S. O'Hern, L. E. Silbert, A. J. Liu, and S. R. Nagel, Jamming at zero temperature and zero applied stress: The epitome of disorder, Phys. Rev. E {\bf 68}, 011306 (2003).

\bibitem{Wyart}M. Wyart, L. E. Silbert, S. R. Nagel, and T. A. Witten, Effects of compression on the vibrational modes of marginally jammed solids, Phys. Rev. E {\bf 72}, 051306 (2005).

\bibitem{Chaudhuri}P. Chaudhuri, L. Berthier, and S. Sastry, Jamming transitions in amorphous packings of frictionless spheres occur over a continuous range of volume fractions, Phys. Rev. Lett. {\bf 104}, 165701 (2010).

\bibitem{Vagberg.PRE.2011}D. V{\aa}gberg, P. Olsson, and S. Teitel, Glassiness, rigidity, and jamming of frictionless soft core disks, Phys. Rev. E {\bf 83}, 031307 (2011).





\bibitem{OT1}P. Olsson and S. Teitel, ``Critical scaling of shear viscosity at the jamming transition," Phys. Rev. Lett. {\bf 99}, 178001 (2007).

\bibitem{OT2}P.~Olsson and S.~Teitel, ``Critical scaling of shearing rheology at the jamming transition of soft-core frictionless disks,"  Phys. Rev. E {\bf 83}, 030302(R) (2011).

\bibitem{VagbergOlssonTeitel}D. V{\aa}gberg, P. Olsson, and S. Teitel, ``Critical scaling of Bagnold rheology at the jamming transition of frictionless two-dimensional disks," Phys. Rev. E {\bf 93}, 052902 (2016).

\bibitem{OT3}P.~Olsson and S.~Teitel, ``Herschel-Bulkley shearing rheology near the athermal jamming transition," Phys. Rev. Lett. {\bf 109}, 108001 (2012).



\bibitem{Hatano1}T. Hatano, Scaling properties of granular rheology near the jamming transition, J. Phys. Soc. Jpn. {\bf 77}, 123002 (2008).

\bibitem{Hatano2}T. Hatano, Growing length and time scales in a suspension of athermal particles, Phys. Rev. E {\bf 79},  050301R (2009).  

\bibitem{Hatano3}T. Hatano, Critical scaling of granular rheology, Prog. Theor. Phys. Suppl. {\bf 184}, 143 (2010).

\bibitem{Otsuki}M. Otsuki and H. Hayakawa, Critical behaviors of sheared frictionless granular materials near the jamming transition, Phys. Rev. E {\bf 80}, 011308 (2009).

\bibitem{Heussinger1}C. Heussinger and J.-L. Barrat, Jamming transition as probed by quasistatic shear flow, Phys. Rev Lett. {\bf 102}, 218303 (2009).

\bibitem{Heussinger2}C. Heussinger, P. Chaudhuri, and J.-L. Barrat, Fluctuations and correlations during the shear flow of elastic particles near
the jamming transition, Soft Matter {\bf 6}, 3050 (2010).





\bibitem{Silbert}L. E. Silbert, A. J. Liu, and S. R. Nagel, Vibrations and diverging length scales near the unjamming transition, Phys. Rev. Lett. {\bf 95}, 098301 (2005).

\bibitem{Wyart2}M. Wyart, S. R. Nagel, and T. A. Witten, Geometric origin of excess low-frequency vibrational modes in weakly connected amorphous solids, Europhys. Lett. {\bf 72}, 486 (2005).

\bibitem{Hexner}D. Hexner, A. J. Liu, and S. R. Nagel, Two diverging length scales in the structure of jammed packings, Phys. Rev. Lett. {\bf 121}, 115501 (2018).

\bibitem{Drocco}J. A. Drocco, M. B. Hastings, C. J. Olson Reichhardt, and C. Reichhardt, Multiscaling at point J: Jamming is a critical phenomenon, Phys. Rev. Lett. {\bf 95}, 088001 (2005).

\bibitem{VOT}D. V{\aa}gberg, D. Valdez-Balderas, M. A. Moore, and P. Olsson and S. Teitel, ``Finite-size scaling at the jamming transition: Corrections to scaling and the correlation-length critical exponent," Phys. Rev. E  {\bf 83}, 030303(R) (2011)


\bibitem{Wyart3}M. Wyart, On the rigidity of amorphous solids, Ann. Phys. Fr. {\bf 30}, 1 (2005). 

\bibitem{Goodrich}C. P. Goodrich, A. J. Liu, and S. R. Nagel, Finite-size scaling at the jamming transition, Phys. Rev. Lett. {\bf 109}, 095704 (2012).

\bibitem{Charbonneau}P. Charbonneau, E. I. Corwin, G. Parisi, and F. Zamponi, Universal microstructure and mechanical stability of jammed packings, Phys. Rev. Lett. {\bf 109}, 205501 (2012).


\bibitem{Goodrich2}C. P. Goodrich, S. Dagois-Bohy, B. P. Tighe, M. van Hecke, A. J. Liu, and S. R. Nagel, Jamming in finite systems: Stability, anisotropy, fluctuations, and scaling, Phys. Rev. E {\bf 90}, 022138 (2014).


\bibitem{Heussinger3}C. Heussinger, L. Berthier, and J.-L. Barrat, Superdiffusive, heterogeneous, and collective particle motion near the fluid-solid transition in athermal disordered materials, Europhys. Lett. {\bf 90}, 20005  (2010).


\bibitem{During2}G. D{\"u}ring, E. Lerner and M. Wyart, Length scales and self-organization in dense suspension flows, Phys. Rev. E {\bf 89}, 022305 (2014).

\bibitem{Pouliquen}O. Pouliquen, Velocity correlations in dense granular flows, Phys. Rev. Lett. {\bf 93}, 248001 (2004).

\bibitem{Durian}D.~J.~Durian, ``Foam mechanics at the bubble scale," Phys. Rev. Lett. {\bf 75}, 4780 (1995) and ``Bubble-scale model of foam mechanics: Melting, nonlinear behavior, and avalanches," Phys. Rev. E {\bf 55}, 1739 (1997).

\bibitem{LeesEdwards}D.~J.~Evans and G.~P.~Morriss, {\em Statistical Mechanics of Non-equilibrium Liquids} (Academic Press, London, 1990).

\bibitem{Tewari}S. Tewari, D. Schiemann, D. J. Durian, C, M. Knobler, S. A. Langer, and A. J. Liu, Statistics of shear-induced rearrangements in a two-dimensional model foam, Phys. Rev. E {\bf 60}, 4385 (1999).

\bibitem{Andreotti}B. Andreotti, J.-L. Barrat, and C. Heussinger, ``Shear flow of non-brownian suspensions close to jamming,"  Phys. Rev. Lett. {\bf 109}, 105901 (2012).

\bibitem{Lerner}E. Lerner, G. D{\"u}ring, and M. Wyart, ``A Unified framework for non-Brownian suspension flows and soft amorphous solids," Proc. Natl. Acd. Sci. U.S.A. {\bf 109}, 4798 (2012).

\bibitem{Vagberg.PRL.2014}D. V{\aa}gberg, P. Olsson, and S. Teitel, ``Universality of jamming criticality in overdamped shear-driven frictionless disks," Phys. Rev. Lett. {\bf 113}, 148002 (2014).

\bibitem{DeGiuli}E. DeGiuli, G. D{\"u}ring, E. Lerner, and M. Wyart, ``Unified theory of inertial granular flows and non-Brownian suspensions," Phys. Rev. E {\bf 91}, 062206 (2015).

\bibitem{Berthier}T. Kawasaki, D. Coslovich, A. Ikeda, and L. Berthier, ``Diverging viscosity and soft granular rheology in non-Brownian suspensions," Phys. Rev. E {\bf  91}, 012203 (2015).

\bibitem{gxvsg}We find that $g(x)$ has a better fit to an exponential decay at smaller $x$ than does $g_x(x)$, and that the length $\xi$ extracted from $g(x)$ is about 1.5 times smaller than that from $g_x(x)$; this means that finite size effects set in for $g_x(x)$  at larger $\dot\gamma$  than they do for $g(x)$.

\bibitem{Chaikin}P. M. Chaikin and T. C. Lubensky, {\em Principles of Condensed Matter Physics}, (Cambridge University Press, Cambridge, 1995), see Chapter 5.

\bibitem{Binder}K. Binder, Finite size scaling analysis of ising model block distribution functions, Z. Phys. B {\bf 43}, 119 (1981).

\bibitem{Hasenbusch}M. Hasenbusch, A. Pelissetto, and E. Vicari, The critical behavior of 3D Ising spin glass models: universality and scaling corrections, J. Stat. Mech. (2008) L02001.

\bibitem{Olsson1}P. Olsson, Relaxation times and rheology in dense athermal suspensions, Phys. Rev. E {\bf 91}, 062209 (2015).

\bibitem{Olsson2}P. Olsson, Asymmetric velocity correlations in shearing media, Phys. Rev. E {\bf 82}, 031303 (2010).














\end{thebibliography}

\end{document}